\documentclass[epj]{svjour}
\usepackage{multicol}
\usepackage{graphics}
\usepackage{hyperref}
\usepackage{titlesec}
\usepackage{graphicx,epsfig,epstopdf}
\usepackage{bm}
\usepackage{epstopdf}
\usepackage{amsmath,amsfonts,latexsym}
\usepackage{dcolumn}
\usepackage{bm}
\usepackage{hyperref}
\usepackage[italic]{hepnames}
\usepackage[utf8]{inputenc}

\usepackage{breqn}
\usepackage{breqn}
\usepackage{xcolor}
\usepackage{booktabs}
\usepackage{bm,mathrsfs,bbm,amscd}
\usepackage{array,multirow,graphicx}
\graphicspath{{Figures/}} %
\begin{document}

\title{Bottomonium spectroscopy using Coulomb plus linear (Cornell) potential}

\author{ Virendrasinh Kher$^{a}$, Raghav Chaturvedi$^b$, Nayneshkumar Devlani $^{a}$, and A. K. Rai$^c$
}                     
\offprints{vhkher@gmail.com\\
raghavr.chaturvedi@gmail.com\\
raiajayk@gmail.com}          
\institute{$^{a}$Applied Physics Department, Polytechnic, The Maharaja Sayajirao University of Baroda, Vadodara 390002,  Gujarat, {\it INDIA}\\
$^{b}$Ministry of Education, Abu Dhabi, {\it UAE}\\
$^{c}$Department of Physics, Sardar Vallabhbhai National Institute of Technology, Surat 395007, Gujarat, {\it INDIA}} 
\date{Received: date / Revised version: date}
%
\abstract{
The coulomb plus linear (Cornell) potential is used to investigate the mass spectrum of bottomonium. Gaussian wave function is used in position and momentum space to estimate values of potential and kinetic energies, respectively. Based on our calculations, we study newly observed ${{\mathit \Upsilon}{(10860)}}$ as an admixture of $4^{3}S_{1}$ with $4^{3}D_{1}$, and ${{\mathit \Upsilon}{(10753)}}$ as an admixture of $6^{3}S_{1}$ with $4^{3}D_{1}$ also, we try to assign ${{\boldsymbol \Upsilon}{(11020)}}$ as a pure $4^{3}D_{1}$ bottomonium state.  We also study the Regge trajectories in the $(J,M^{2})$ and $(n_r,M^{2})$ planes to help prove our association. We estimate the pseudoscalar and vector decay constants, the radiative (Electric and Magnetic Dipole) transition rates, and the annihilation decay width for bottomonium states.}

\authorrunning { V Kher, R Chaturvedi, N. Devlani, A K Rai}
\titlerunning {Bottomonium spectroscopy using Coulomb plus linear (Cornell) potential}
\maketitle

\section{Introduction}
\label{intro}
Bottomonium ($b\overline{b}$ meson) was discovered as $\Upsilon$, $\Upsilon^{\prime}$ and $\Upsilon^{\prime\prime}$ by the E288 Collaboration at Fermilab in 1977, and were substantially studied at various $e^{+}e^{-}$ storage rings\cite{Zyla:2020zbs,Innes:1977}. In 1982,  $\chi_{bJ}$ $(2P)$ states and in 1983, $\chi_{bJ}$ $(2P)$, $(J=1,2,3)$ states were discovered in E1 transition from $\Upsilon^{\prime\prime}$ \cite{Han:1982,Eigen:1982} and  $\Upsilon^{\prime}$  \cite{Klopfenstein:1983,Pauss:1983}, respectively. In 1984, $\Upsilon (4S)$,  $\Upsilon (5S)$ and  $\Upsilon (6S)$ states were observed \cite{Besson:1984,Lovelock:1985}.
 In 2008, after three decade of discovery of $\Upsilon (nS)$ resonance, the BaBar Collaboration found  $\eta_{b}$, the pseudoscalar partner of the
 triplet state $\Upsilon (1S)$\cite{Aubert2008}. In 2012, the Belle collaboration reported the first evidence of the $\eta _b (2S)$ with mass $9999\pm 3.5_{-1.9}^{+2.8}$,$MeV/c^2$ using $h_b(2P) \rightarrow \gamma \eta_b (2S)$ transition and first observation of $h_b(1P) \rightarrow \gamma \eta_b (1S)$ and $h_b(2P) \rightarrow \gamma \eta_b (1S)$  with mass  $9402 \pm 1.5\pm 1.8$ $MeV/c^2$ \cite{Mizuk:2012}.

 In 2004, CLEO Collaboration presented the first evidence for the production of  $\Upsilon (1S)$(1D) states in the four-photon cascade, such as in a two-photon cascade starting from the $\Upsilon(3S) \rightarrow \gamma\chi_{b}(2P_{J})$, $\chi_{b}(2P_{J}) \rightarrow \gamma\Upsilon(1D)$ and then select events with two more subsequent photon transitions, $\Upsilon(1D) \rightarrow \gamma\chi_{b}(1P_{J})$, $\chi_{b}(1P_{J}) \rightarrow \gamma\Upsilon(1S)$, followed by the $\Upsilon(1S)$ annihilation into either $e^{+}e^{-}$ or $\mu^{+}\mu^{-}$\cite{Bonvicini:2004}. In 2010, the BABAR Collaboration reported the observation of the $J = 2$
 state of $\Upsilon(1^{3}D_{J})$ in the hadronic $\pi^{+}\pi^{-}\Upsilon(1S)$ decay channel, with
 $\Upsilon(1S)\rightarrow e^{+}e^{-}\,\text{or}\, \mu^{+}\mu^{-}$\cite{delAmoSanchez:2010}. In 2011, BABAR Collaboration reported evidence for the $h_{b}(1P)$ state
 in the decay $\Upsilon(3S) \rightarrow \pi{0}h_b(1P) $, with data sample corresponds to 28 $fb^{-1}$ of integrated luminosity at a center of mass energy of 10.355 GeV, the mass of the $\Upsilon(3S)$ resonance and in the same year, the Belle Collaboration reported the first observation of the spin-singlet bottomonium states $h_b(1P)$ and $h_b(2P)$ produced via $e^{+}e^{-}\rightarrow h_b(nP)\pi^{+}\pi^{-}$ transition corresponds to 121.4 $fb^{-1}$ of integrated luminosity  near the peak of the $\Upsilon(5S)$ resonance at a center-of-mass energy $\sqrt{s}\sim 10.865$ GeV\cite{Adachi:2011}. Again in 2011, ATLAS Collaboration observed the $\chi_{b}(nP)$ states, recorded by the ATLAS detector during the proton-proton collisions at the LHC, run at a center-of-mass energy $\sqrt{s}\sim 7\,TeV$ and these states were reconstructed through their radiative decays to $\Upsilon (1S, 2S)$ with $\Upsilon \rightarrow \mu^{+}\mu^{-}$. In addition to the mass
 peaks corresponding to the $\chi_{b}(1P,2P) \rightarrow \gamma\Upsilon(1S,2S)$, at a mass of $10.530\pm0.005(stat.)\pm0.009(syst.)$ GeV in both the decay modes and structure has been assigned to the $\chi_{b}(3P)$ multiplet\cite{Aad:2011}.

 Recently in 2014, the LHCb Collaboration has determined the mass of the $\chi_{b1}(3P)$ using its radiative decays to the $\Upsilon (1S)$ and $\Upsilon (1S)$ mesons and the measured mass of the meson is $m(\chi_{b1}(3P))=10515.7_{-3.9}^{+2.2}(stat.)_{-2.1}^{+1.5}(syst.)$ $MeV/c^2$\cite{Aaij:2014b}. Many other states above open flavour have also been observed experimentally in the Bottomonium family but their association to a particular S,P or D state remains questionable. It is because of unknown couplings of the many channels that become available once the threshold is crossed. Two charged states namely ${{\boldsymbol Z}_{{b}}{(10610)}}$ and ${{\mathit Z}_{{b}}{(10650)}}$ were observed at Belle by \cite{Belle:2011aa} in 2011 in the ${{\mathit \Upsilon}{(5S)}}$ decays to ${{\mathit \Upsilon}{(nS)}}{{\mathit \pi}^{+}}{{\mathit \pi}^{-}}$ (n = 1, 2, 3) and ${{\mathit h}_{{b}}{(mP)}}{{\mathit \pi}^{+}}{{\mathit \pi}^{-}}$ (m = 1, 2) with $\mathit J{}^{P} = 1{}^{+}$. Ref. \cite{Bondar:2011ev} suggested by considering molecular nature of these states all their properties can be explained. Recently, in 2019 one new state ${{\boldsymbol \Upsilon}{(10753)}}$ was observed at Belle by \cite{Abdesselam:2019gth} in ${{\mathit e}^{+}}{{\mathit e}^{-}}\rightarrow {{\mathit \Upsilon}{(nS)}}{{\mathit \pi}^{+}}{{\mathit \pi}^{-}}$  (n=1,2,3) with a significance of 5.2 $\sigma$ with mass of $10753 \pm6$ MeV and width of $36 {}^{+18}_{-12}$ MeV. This new observed state can be a candidate for ${{\mathit \Upsilon}{(3D)}}$ state as suggested by \cite{Chen:2019uzm}, or a compact tetraquark \cite{Ali:2009es} or hadrobottomonium \cite{Alberti:2016dru}.

 Lately, there has been renewed interest in the study of highest bottomonium states ${{\Upsilon}{(10860)}}$ and ${{\mathit \Upsilon}{(11020)}}$ which were first observed in 1985 at CUSB by \cite{Lovelock:1985nb}. Fresh study of these two states was conducted at BELL in 2019 by \cite{Abdesselam:2019gth}, the calculated masses found are $10885.3 \pm1.5 {}^{+2.2}_{-0.9}$ and $11000.0 {}^{+4.0}_{-4.5} {}^{+1.0}_{-1.3}$ MeV, and the decay widths found are $36.6  {}^{+4.5}_{-3.9} {}^{+0.5}_{-1.1}$ and $23.8  {}^{+8.0}_{-6.8} {}^{+0.7}_{-1.8}$ MeV. Ref. \cite{Chen:2019uzm,Molina:2020zao} suggest ${{\Upsilon}{(10860)}}$ as $5^{3}S_1$ bottomonium state, while \cite{Pandya:2021} and \cite{Chaturvedi:2020} suggests it as an admixture of $5^{3}S_1$ and $6^{3}D_1$, and $5^{3}S_1$ and $5^{3}D_1$ states. ${{\mathit \Upsilon}{(11020)}}$ has been suggested as an admixture of $6^{3}S_1$ and $5^{3}D_1$ by \cite{Chaturvedi:2020,Pandya:2021}, also it has been well reproduced by other potential models that take into account coupled channel effects \cite{Liu:2011yp}. Further, the next generation B-factory experiment Belle II at SuperKEKB collider with luminosity approximately 50 times greater than Belle I will offer promising prospects for bottomonium physics.\\
 These experimental exploration in  the observation of new bottomonium states, both conventional and unconventional  has re-ignited interest and motivated many theoreticians to carry out a comprehensive study. Comprehensive theoretical and experimental literature can be found in \cite{Eichten:2008,Brambilla:2011,Brambilla:2014}, also certain methods like lattice QCD \cite{McNeile:2012qf}, heavy quark effective field theory (EFT) \cite{Neubert:1993mb}, chiral perturbation theory \cite{Pentia:2009zz}, dynamical equation-based approaches like the Schwinger–Dyson and Bethe–Salpeter equations \cite{Ricken:2000kf,Mitra:1990av}, QCD sum rules \cite{Otsuka:2018bqq}, NRQCD \cite{Mateu:2018zym}, non-relativistic EFT \cite{Brambilla:2019jfi}, an effective super-symmetric approach \cite{Nielsen:2018uyn}, quark models \cite{Li:2019tbn},  QCD-motivated relativistic quark model based on the quasipotential approach \cite{Ebert:2011jc}, and various potential models have tried to compute the spectroscopic properties of bottomonium. \\
Theoretical studies will allow one to single out experimental candidates and prove to be powerful tools for understanding the quark–antiquark interaction as expected from quantum chromodynamics (QCD). \textbf{From a theoretical viewpoint, non relativistic potential models; where quark antiquark interaction is modeled by a potential function have been very successful to the study of heavy quarkonium. Considering the complex structure of the QCD vacuum, it is difficult to obtain 	quark–antiquark interaction potential starting from basic principles of QCD. One of the earliest potential functions to be employed is the Cornell potential \cite{Eichten1978,Eichten1980}. In the absence of closed-form solutions for the Schr\"{o}dinger equation with Cornell potential one has to resort to various approximate methods or numerical solutions\cite{Lucha1999}. In literature we find many studies of quarkonium with Cornell potential. Previously Hall \cite{Hall1984} has employed a simple eigenvalue formula under some restrictions to obtain energy eigenvalues, Vega et.al. \cite{Vega2016} have employed variational method in the light of Supersymmetric Quantum Mechanics, Jacobs et. al. \cite{Jacobs1986} develop a unified approach to the solution of the Schrödinger and the spinless Salpeter equations and obtained the eigenvalues.} \textbf{In the present article we employ variational method with a single Gaussian trial wave function with one parameter; both in position space as well as momentum space; in a Cornell potential model to calculate the mass spectrum of the $b\bar{b}$ meson. Such a wavefunction is quite accurate in the long range part of the potential however it has a limitation that in the short range part the deviations can turn larger as $r \to 0$. In this regard a multi-gaussian approach could be investigated. However, the present approach offers the advantage that it gives an analytical expression for the wavefunction that has a simple form. The wavefunction can then be conveniently employed for physical application.} We incorporate corrections to the kinetic energy of quarks as well as the relativistic correction of ${\cal{O}}\left(\frac{1}{m}\right)$ to the potential energy part of the Hamiltonian. Using our predicted masses for the $b\bar{b}$ meson, we  also plot the Regge trajectories in both the $(M^{2}\rightarrow J)$ and $(M^{2}\rightarrow n)$ planes (where $J$ is the spin and $n$ is the principal quantum number), as the Regge trajectories play a important role to identify the nature of current and future experimentally observed $b\bar{b}$ meson. We also estimate the pseudoscalar and vector decay constants, the radiative (Electric and Magnetic Dipole) transition rates between D, P and S wave as well as the annihilation decays for bottomonium states.

 The article is organized as follows. We present the theoretical framework for the mass spectra in Section {\ref{sec:mass}}, the decay constants ($f_{P/V}$) in Section {\ref{sec:decay}}, the radiative (E1 and M1) transitions in Section {\ref{sec:E1M1}} and annihilation decays in Section {\ref{sec:annihilation}}.  Results for the  mass spectra, ($f_{P/V}$) decays, E1 and M1 transition width as well as annihilation decays for the $b\bar{b}$ meson, are discussed in Section {\ref{sec:Resu}}. The Regge trajectories in the $(J,M^{2})$ and $(n_r,M^{2})$ planes are in Section {\ref{sec:reg}}. Finally, we draw our conclusion in  Section {\ref{sec:conclusion}}.

\section{Method}

\subsection{Cornell potential with ${\cal{O}}\left(\frac{1}{m}\right)$ corrections \label{sec:mass}}
Here we compute the mass spectra of bottomonium using the  Coulomb plus linear potential, the Cornell potential \cite{Eichten1978,Eichten1980}. Here, we incorporate relativistic approach by adding relativistic corrections to the kinetic energy part and ${\cal{O}}\left(\frac{1}{m}\right)$ correction to the potential energy part \cite{kher2018,Koma2006}. We employ the Cornell potential for bottomonium, as it works well for charmonium, proven by our previous work~\cite{kher2018}. Here, we employ the following Hamiltonian \cite{Gupta1995,Hwang1997,Devlani2014,Kher2017b}  and quark-antiquark potential \cite{Eichten1978,Koma2006} ,

\begin{equation}
H=\sqrt{\mathbf{p}^{2}+m_{Q}^{2}}+\sqrt{\mathbf{p}^{2}+m_{\bar{Q}}^{2}}+V(\mathbf{r}),\label{Eq:hamiltonian}
\end{equation}
\begin{equation}
V\left(r\right)=V^{\left(0\right)}\left(r\right)+\left(\frac{1}{m_{Q}}+\frac{1}{m_{\bar{Q}}}\right)V^{\left(1\right)}\left(r\right)+{\cal O}\left(\frac{1}{m_{}^{2}}\right).
\end{equation}
The Cornell-like potential $V^{\left(0\right)}$ \cite{Eichten1978} and $V^{\left(1\right)}\left(r\right)$ from leading order perturbation theory  are\cite{Koma2006},
\begin{equation}\label{pote}
V^{\left(0\right)}(r)=-\frac{4\alpha_{S}\left({M^{2}}\right)}{3r}+Ar+V_{0}
\end{equation}
\begin{equation}
V^{\left(1\right)}\left(r\right)=-C_{F}C_{A}\alpha_{s}^{2}/4r^{2}
\end{equation}
where,  $m_{Q}$ is the quark and $m_{\bar{Q}}$ anti-quark mass, $V_{0}$ is the potential constant,  $A$ is the potential parameter,
$\alpha_{S}\left({M^{2}}\right)$ is the strong running coupling constant  and $C_{F}=4/3$, $C_{A}=3$ are the Casimir charges \cite{Koma2006}.

To calculate the expectation values of the Hamiltonian with the Ritz variational strategy, we use a Gaussian wave function\cite{Kher2017b,Kher2017c,kher2018}, which has the form
\begin{eqnarray}
R_{nl}(\mu,r) & = & \mu^{3/2}\left(\frac{2\left(n-1\right)!}{\Gamma\left(n+l+1/2\right)}\right)^{1/2}\left(\mu r\right)^{l}\times\nonumber \\
 &  & e^{-\mu^{2}r^{2}/2}L_{n-1}^{l+1/2}(\mu^{2}r^{2})
\end{eqnarray} and
\begin{eqnarray}
R_{nl}(\mu,p) & = & \frac{\left(-1\right)^{n}}{\mu^{3/2}}\left(\frac{2\left(n-1\right)!}{\Gamma\left(n+l+1/2\right)}\right)^{1/2}\left(\frac{p}{\mu}\right)^{l}\times\nonumber \\
 &  & e^{-{p}^{2}/2\mu^{2}}L_{n-1}^{l+1/2}\left(\frac{p^{2}}{\mu^{2}}\right)
\end{eqnarray} respectively with the Laguerre polynomial $L$ and the variational parameter $\mu$. We estimated $\mu$ for each state, for the preferred value of $A$, using \cite{Hwang1997},
\begin{equation}
\left\langle{K.E.}\right\rangle =\frac{1}{2} \left\langle{\frac{rdV}{dr}}\right\rangle\label{Eq:virial theorem}
\end{equation}

We expand the Hamiltonian (\ref{Eq:hamiltonian}), integrate the relativistic correction and obtain the expectation values of the potential energy as well as kinetic energy as per our previous work on charmonium and other mesons\cite{Kher2017b,Kher2017c,kher2018}.

We have fitted the ground state center of the weight mass  using the Eq.~(\ref{Eq:rai1-1}), by fixing the potential parameters $A$, $\alpha_s$,  $V_0$ and equated with the PDG data as well as we forecast  the center of weight mass for the $nJ$ state using the Eq.~(\ref{Eq:rai2-1}) \cite{Rai2008,kher2018}.

\begin{equation}
M_{SA}=M_{P}+\frac{3}{4}(M_{V}-M_{P}),\label{Eq:rai1-1}
\end{equation}

\begin{equation}
M_{CW,n}=\frac{\Sigma_{J}(2J+1)M_{nJ}}{\Sigma_{J}(2J+1)}\label{Eq:rai2-1}
\end{equation}

The spin dependent part of the potential for computing the mass difference between different degenerate bottomonium states can be written as  \cite{Barnes:2005,Eichten:2008,Voloshin:2007,Lakhina2006,kher2018}.
\begin{eqnarray}
V_{SD} & = & V_{LS}(r)\left(\vec{L}\cdot\vec{S}\right)+ V_{SS}(r)\left[S\left(S+1\right)-\frac{3}{2}\right]+\nonumber\\
 & &  V_{T}(r)\left[S\left(S+1\right)-\frac{3\left(\vec{S}\cdot\vec{r}\right)\left(\vec{S}\cdot\vec{r}\right)}{r^{2}}\right]
\end{eqnarray}
where the spin-spin, spin-orbit and tensor interactions can be written in terms of the vector and scalar parts of $V(r)$ as \cite{Voloshin:2007}
\begin{eqnarray}
V_{SS}(r) & = & \frac{1}{3m_{Q}^{2}}\nabla^{2}V_{V} =\frac{16\pi\alpha_{s}}{9m_{Q}^{2}}\delta^{3}\left(\vec{r}\right),
\end{eqnarray}
\begin{eqnarray}
V_{LS}(r) & = & \frac{1}{2m_{Q}^{2}r}\left(3\frac{dV_{V}}{dr}-\frac{dV_{S}}{dr}\right),
\end{eqnarray}
\begin{eqnarray}
V_{T}(r) & = & \frac{1}{6m_{Q}^{2}}\left(3\frac{d^{2}V_{V}}{dr^{2}}-\frac{1}{r}\frac{dV_{V}}{dr}\right),
\end{eqnarray}
\noindent where
$V_{V}(=-\frac{4\alpha_{s}}{3r})$ is the Coulomb part and $V_{S}( = Ar )$ is the confining part of Eq.(\ref{pote})

In the present study, the quark masses is taken as $m_{b}=4.88$ ~GeV  to reproduce the ground state masses of the bottomonium states. The fitted potential parameters are $A=0.222~ GeV^{2}$, $\alpha_s=0.270$ and $V_0= -0.36923~ GeV$.

\subsection{Mixing in bottomonium states}

There are ambiguities in the structure of many of the heavy quarkonium states. The observed states like $\Upsilon \left(4S\right)\left(10860\right)$, $\Upsilon \left(11020\right)$ are assigned $5^{3}S_{1}$, $4^{3}D_{1}$ in the present work, however there is also a possibility that they could be an admixture of S-D states \cite{Badalian2010}.

Mass of a mixed state $\left(M_{nL}\right)$ is expressed in terms
of  two pure states $\left(nl\,\text{ and }\,n^{\prime}l^{\prime}\right)$
as\cite{Badalian2010,Shah2012}
\begin{equation}
	M_{nL}=\left|\cos^{2}\theta\right|M_{nl}+\left(1-\left|\cos^{2}\theta\right|\right)M_{n^{\prime}l^{\prime}};
\end{equation}
here, $\theta$ is the mixing angle\cite{Radford2011,Shah2012}.

The wave function at origin for D-wave states is defined as\cite{Badalian2010,Shah2012}
\begin{equation}
	R_{D}\left(0\right)=\frac{5R_{D}^{\prime\prime}\left(0\right)}{2\sqrt{2}\omega_{b}^{2}};
\end{equation}
where$R_{D}^{\prime\prime}\left(0\right)$ is the second order derivative
of the wave function at origin for D state and $\omega_{b}$ is a
constant having value $5.11$ GeV. The computed states are listed in table \ref{tab:bbmixing}.



\subsection{Decay Constant ($f_{P/V}$)}
\label{sec:decay}
We calculate decay constant with the QCD correction factor for bottomonium  using the Van-Royen-Weisskopf formula \cite{VanRoyen1967,Braaten1995},
\begin{equation}\label{Eq:decayconst}
f_{P/V}^{2}=\frac{12\left|\psi_{P/V}(0)\right|^{2}}{M_{P/V}}\left(1-\frac{\alpha_{S}}{\pi}\left[2-\frac{m_{Q}-m_{\bar{q}}}{m_{Q}+m_{\bar{q}}}\ln\frac{m_{Q}}{m_{\bar{q}}}\right]\right);\end{equation}
Equation(\ref{Eq:decayconst}) also gives the inequality~\cite{Hwang1997a}
\begin{equation}\label{eq:ineq}
\sqrt{m_v}f_v \geq \sqrt{m_p}f_p
\end{equation}

\subsection{Radiative Transitions \label{sec:E1M1}}
The matrix element of the EM current between the initial ($i$) and final ($f$) quarkonium state, i.e., $\langle f\mid j_{em}^{\mu}\mid i\rangle$, affect the radiative transition. The electric and magnetic dipole transitions are leading order transition amplitudes \cite{Ding:2007,Lu2016,Guo:2010a,Devlani2014,kher2018}.

The E1 matrix elements are estimated by~\cite{Radford2009}
\noindent
\begin{equation}
	\Gamma_{(E1)}\left(n^{2S+1}L_{J}\rightarrow n^{'2S^{'}+1}L_{J^{'}}^{'}+\gamma\right)=
	\frac{4\alpha e_{Q}^{2}}{3} \frac{E_{\gamma}^{3}E_{f}}{M_{i}}C_{fi}\delta_{SS^{'}}\times\left|\left\langle f\left|r\right|i\right\rangle \right|^{2}
\end{equation}
where photon energy  $ E_{\gamma}=\frac{M_{i}^{2}-M_{f}^{2}}{2M_{i}}$,  $\alpha=1/137$, the fine structure constant, $e_{Q}$ is the quark charge,  and $E_f$ is the energy of the final state. The angular momentum matrix element ($C_{fi}$) is
\begin{equation}
C_{fi}=max\left(L,L^{'}\right)\left(2J^{'}+1\right)\left\{ \begin{array}{ccc}
L^{'} & J^{'} & S\\
J & L & 1
\end{array}\right\}^{2}
\end{equation}
where $\left\{ :::\right\} $ is a 6-j symbol. The matrix elements $ \langle n^{'2S^{'}+1}L_{J^{'}}^{'}\mid r\mid n^{2S+1}L_{J}\rangle $ are evaluated using the wave-functions
\begin{equation}
\left\langle f \left| r \right| i \right\rangle=\int dr R_{n_{i}l_{i}}\left(r\right)R_{n_{f}l_{f}}\left(R\right)
\end{equation}

The magnetic dipole M1 rate for transitions between  $S$-wave levels is given by \cite{Li2011,Bardeen2003,Brambilla:2011}
\begin{equation}\label{eq:m1}
 \Gamma_{M1}(i\rightarrow f+\gamma)=\frac{16\alpha}{3}\mu^{2}k^{3}(2J_{f}+1)\left|\left\langle f\right|j_{0}(kr/2)\left|i\right\rangle \right|^{2}\end{equation}

where the magnetic dipole moment is \[
\mu=\frac{m_{\bar{q}}e_{Q}-m_{Q}e_{\bar{q}}}{4m_{\bar{q}}m_{Q}}\].

\subsection{Annihilation Decays \label{sec:annihilation}}

Annihilation decays of quarkonia states are highly useful for  identification and the production of resonances as well as to recognize conventional mesons and multi-quark structures \cite{Kwong:1987,Kwong:1988}.\\

\subsubsection{Leptonic decays}
The $^{3}S_{1}$ and $^{3}D_{1}$ states annihilate into lepton pairs through a single virtual photon ($Q\bar{Q}\rightarrow l^{+}l^{-},\,\,\text{where}\,\,l=e^{-},\,{\mu}^{-},\,\tau^{-}$). The leptonic decay width of the ($^{3}S_{1}$) and ($^{3}D_{1}$) states of bottomonium, including first order radiative QCD correction, is given by \cite{Segovia2016,Kwong:1987,Bradley:1980}:

\begin{equation}
\varGamma\left(n^{3}S_{1}\rightarrow e^{+}e^{-}\right)=\frac{4e_{Q}^{4}\alpha^{2}\mid R_{nS}\left(0\right)\mid^{2}}{M_{nS}^{2}}\left(1-\frac{16\alpha_{s}}{3\pi}\right)
\end{equation}
\begin{equation}
\varGamma\left(n^{3}D_{1}\rightarrow e^{+}e^{-}\right)=\frac{25e_{Q}^{2}\alpha^{2}\mid R_{nD}^{\prime\prime}\left(0\right)\mid^{2}}{2m_{Q}^{4}M_{nD}^{2}}\left(1-\frac{16\alpha_{s}}{3\pi}\right)
\end{equation}
where, $M_{nS}$ is the mass of the decaying $b\bar{b}$ meson state.\\

\subsubsection{Decay into photons }
The annihilation decay of the bottomonium states into di-photons or tri-photons, with and/or without radiative QCD corrections are given by~\cite{Segovia2016,Kwong:1987}:

\begin{equation}
\varGamma\left(n^{1}S_{0}\rightarrow\gamma\gamma\right)=\frac{3e_{Q}^{4}\alpha^{2}\mid R_{nS}\left(0\right)\mid^{2}}{m_{Q}^{2}}\left(1-\frac{3.4\alpha_{s}}{\pi}\right)
\end{equation}
\begin{equation}
\varGamma\left(n^{3}P_{0}\rightarrow\gamma\gamma\right)=\frac{27e_{Q}^{4}\alpha^{2}\mid R_{nP}^{\prime}\left(0\right)\mid^{2}}{m_{Q}^{4}}\left(1+\frac{0.2\alpha_{s}}{\pi}\right)
\end{equation}
\begin{equation}
\varGamma\left(n^{3}P_{2}\rightarrow\gamma\gamma\right)=\frac{36e_{Q}^{4}\alpha^{2}\mid R_{nP}^{\prime}\left(0\right)\mid^{2}}{5m_{Q}^{4}}\left(1-\frac{16\alpha_{s}}{3\pi}\right)
\end{equation}
\begin{equation}
\varGamma\left(n^{3}S_{1}\rightarrow3\gamma\right)=\frac{4 (\pi^{2}-9)e_{Q}^{6}\alpha^{3}\mid R_{nS}\left(0\right)\mid^{2}}{3\pi m_{Q}^{2}}\left(1-\frac{12.6\alpha_{s}}{\pi}\right)
\end{equation}\\

\subsubsection{Decay into gluons }
The annihilation decay of the bottomonium states into di-gluons or tri-gluons and  gluons with photons and light quarks, with and/or without radiative QCD correction, are given by~\cite{Segovia2016,Kwong:1987,Kwong:1988,Belanger:1987}:

\begin{equation}
\varGamma\left(n^{1}S_{0}\rightarrow gg\right)=\frac{2\alpha_{s}^{2}\mid R_{nS}\left(0\right)\mid^{2}}{3m_{Q}^{2}}\left(1+\frac{4.4\alpha_{s}}{\pi}\right)
\end{equation}
\begin{equation}
\varGamma\left(n^{3}P_{0}\rightarrow gg\right)=\frac{6\alpha_{s}^{2}\mid R_{nP}^{\prime}
\left(0\right)\mid^{2}}{m_{Q}^{4}}
\end{equation}
\begin{equation}
\varGamma\left(n^{3}P_{2}\rightarrow gg\right)=\frac{8\alpha_{s}^{2}\mid R_{nP}^{\prime}
\left(0\right)\mid^{2}}{5m_{Q}^{4}}
\end{equation}
\begin{equation}
\varGamma\left(n^{1}D_{2}\rightarrow gg\right)=\frac{2\alpha_{s}^{2}\mid R_{nD}^{\prime\prime}
\left(0\right)\mid^{2}}{3\pi m_{Q}^{6}}
\end{equation}
\begin{equation}
\varGamma\left(n^{3}S_{1}\rightarrow3g\right)=\frac{10 (\pi^{2}-9)\alpha_{s}^{3}\mid R_{nS}\left(0\right)\mid^{2}}{81\pi m_{Q}^{2}}\left(1-\frac{4.9\alpha_{s}}{\pi}\right)
\end{equation}
\begin{equation}
\varGamma\left(n^{1}P_{1}\rightarrow3g\right)=\frac{20 \alpha_{s}^{3}\mid R_{nP}^{\prime}\left(0\right)\mid^{2}}{9\pi m_{Q}^{4}}\ln(m_{Q}\langle r \rangle)
\end{equation}
\begin{equation}
\varGamma\left(n^{3}D_{1}\rightarrow3g\right)=\frac{760 \alpha_{s}^{3}\mid R_{nP}^{\prime\prime}\left(0\right)\mid^{2}}{81\pi m_{Q}^{6}}\ln(4m_{Q}\langle r \rangle)
\end{equation}
\begin{equation}
\varGamma\left(n^{3}D_{2}\rightarrow3g\right)=\frac{10 \alpha_{s}^{3}\mid R_{nP}^{\prime\prime}\left(0\right)\mid^{2}}{9\pi m_{Q}^{6}}\ln(4m_{Q}\langle r \rangle)
\end{equation}
\begin{equation}
\varGamma\left(n^{3}D_{3}\rightarrow3g\right)=\frac{40 \alpha_{s}^{3}\mid R_{nP}^{\prime\prime}\left(0\right)\mid^{2}}{9\pi m_{Q}^{6}}\ln(4m_{Q}\langle r \rangle)
\end{equation}
\begin{equation}
\varGamma\left(n^{3}S_{1}\rightarrow \gamma gg\right)=\frac{8 (\pi^{2}-9)e_{Q}^{2}\alpha\alpha_{s}^{2}\mid R_{nS}\left(0\right)\mid^{2}}{9\pi m_{Q}^{2}}\left(1-\frac{7.4\alpha_{s}}{\pi}\right)
\end{equation}
\begin{equation}
\varGamma\left(n^{3}P_{1}\rightarrow q\bar{q}+g\right)=\frac{8\eta_{f} \alpha_{s}^{3}\mid R_{nP}^{\prime}\left(0\right)\mid^{2}}{9\pi m_{Q}^{4}}\ln(m_{Q}\langle r \rangle)
\end{equation}


    \begin{figure}
        \centering
       \includegraphics[bb=30bp 60bp 750bp 550bp,clip,width=0.80\textwidth]{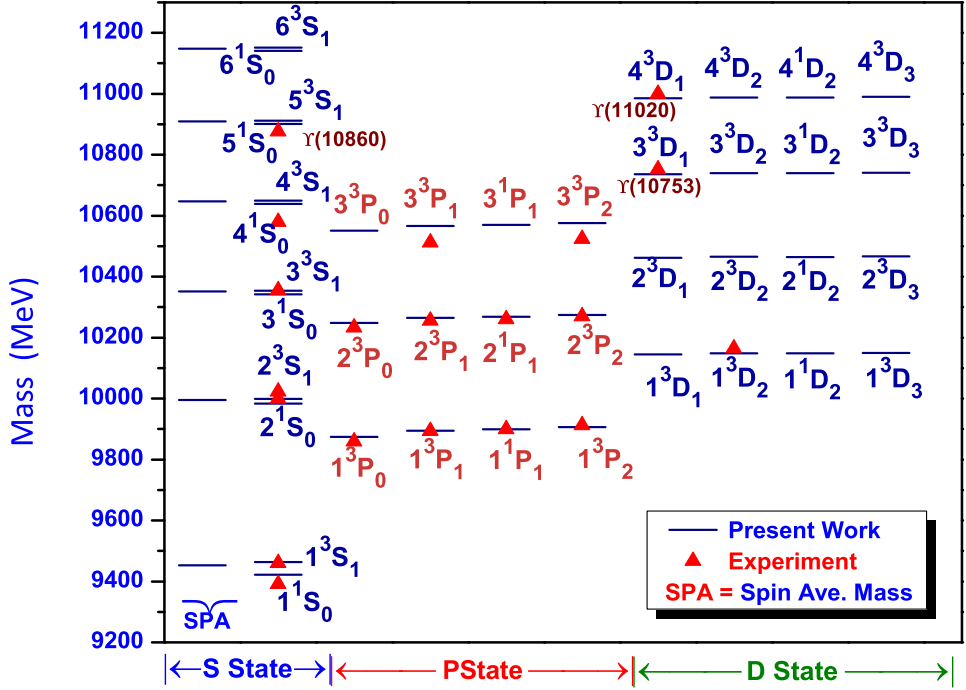}
       \caption{Mass spectrum.\label{fig:MassBB}}
       \end{figure}

  \begin{table*}
      \centering
  \caption{S-P-D-wave center of weight masses  (in GeV).\label{tab:bbmsa}}
    \scalebox{0.95}{
   \begin{tabular}{cccccccccccc}
    \addlinespace[5pt]
       \hline
  \addlinespace[5pt]
   \multirow{2}{*}{$nL$} & \multicolumn{2}{c}{This work} & \multirow{2}{*}{Expt. \cite{Zyla:2020zbs}} & \multicolumn{8}{c}{Others Theory $M_{CW}$  } \tabularnewline
     \cline{2-3} \cline{5-12}
     \addlinespace[3pt]
     & $\mu$ &  $M_{CW}$ &   &Ref.~\cite{Wang:2018}& Ref.~\cite{Bhat:2017} & Ref.~\cite{Deng:2017} & Ref.~\cite{Lu:2016mbb} & Ref.~\cite{Segovia2016} & Ref.~\cite{Godfrey:2015}& Ref.~\cite{Wurtz:2015}&    Ref.~\cite{Li:2009nr}  \tabularnewline

    \addlinespace[2pt]
    \hline
  \addlinespace[3pt]
      $1S$ & 1.216 & 9.453 & 9.453 &9.447 &9.446 & 9.443 & 9.443 & 9.490 & 9.449 & 9.446 & 9.442\tabularnewline

      $2S$ & 0.768 &  9.995 &  &10.010 &10.018 & 10.009 & 10.004 & 10.009 & 9.996 & 10.015 & 10.009\tabularnewline

      $3S$ & 0.664 &  10.351 &  &10.351& 10.388 & 10.339 & 10.368 & 10.344 & 10.350 & 10.329 & 10.346\tabularnewline

      $4S$ & 0.613 &  10.647 &  &10.608& 10.702 & 10.594 &  &  & 10.632 &  & 10.607\tabularnewline

      $5S$ & 0.581 &  10.909 &  &10.819& 10.989 & 10.808 &  &  & 10.876 &  & 10.828\tabularnewline

      $6S$ & 0.557 &  11.148 &  &10.998& 11.295 & 10.995 &  &  & 11.101 &  & 11.020\tabularnewline
   \addlinespace
      $1P$ & 0.811 &  9.899 & 9.899 &9.896 &9.905 & 9.909 & 9.890 & 9.879 & 9.884 & 9.901 & 9.905\tabularnewline

      $2P$ & 0.678 &  10.268 & 10.260 & 10.260&10.276 & 10.254 & 10.263 & 10.240 & 10.252 & 10.219 & 10.258\tabularnewline

      $3P$ & 0.620 &  10.570 &  & 10.531&10.585 & 10.519 & 10.560 &  & 10.542 &  & 10.530\tabularnewline
    \addlinespace
      $1D$ & 0.719 &  10.149 &  & 10.149&10.164 & 10.154 &  & 10.123 & 10.149 & 10.164 & 10.152\tabularnewline

      $2D$ & 0.641 &  10.465 &  &10.465 &10.480 & 10.433 &  & 10.419 & 10.450 & 10.441 & 10.439\tabularnewline

      $3D$ & 0.598 &  10.740 &  &10.740 &10.767 &  &  &  & 10.706 &  & 10.677\tabularnewline
      $4D$ & 0.570 &  10.990 &  &10.988 &11.036 &  & & & &    &        \tabularnewline
    \hline
      \end{tabular}}
      \end{table*}

  \begin{table*}
  	\begin{centering}
  		\caption{Complete mass spectra(in GeV).\label{tab:massesbb}}
  		\par\end{centering}
  	\centering{}\scalebox{0.95}{ %
  		\begin{tabular}{llllllllllll}
  			\toprule
  			\addlinespace
  			 \multirow{2}{*}{State}  & \multirow{1}{*}{$J^{P}$} & This  & Expt.  & \multicolumn{6}{c}{Others} &  &  \tabularnewline \cmidrule{5-12}
  			\addlinespace
  			  &  & work  & Ref.~\cite{Zyla:2020zbs}  & Ref.~\cite{Wang:2018}  & Ref.~\cite{Chen:2019uzm} & Ref.~\cite{Deng:2017}  & Ref.~\cite{Lu:2016mbb}  & Ref.~\cite{Segovia2016}  & Ref.~\cite{Godfrey:2015}  & Ref.~\cite{Wurtz:2015}  & Ref.~\cite{Li:2009nr} \tabularnewline
  			\addlinespace
  			\midrule
  			$1^{1}S_{0}$  & $0^{-+}$  & 9.423  & 9.399  & 9.398 &  & 9.390  & 9.395  & 9.455  & 9.402  & 9.402  & 9.389 \tabularnewline
  			$1^{3}S_{1}$  & $1^{--}$  & 9.463  & 9.460  & 9.463 &  & 9.460  & 9.459  & 9.502  & 9.465  & 9.460  & 9.460 \tabularnewline
  			$2^{1}S_{0}$  & $0^{-+}$  & 9.983  & 9.999  & 9.989 & 9.999 & 9.990  & 9.982  & 9.990  & 9.976  & 9.998  & 9.987 \tabularnewline
  			$2^{3}S_{1}$  & $1^{--}$  & 10.001  & 10.023  & 10.017 & 10.023 & 10.015  & 10.011  & 10.015  & 10.003  & 10.020  & 10.016\tabularnewline
  			$3^{1}S_{0}$  & $0^{-+}$  & 10.342  &  & 10.336 & 10.337 & 10.326  & 10.353  & 10.330  & 10.336  & 10.314  & 10.330 \tabularnewline
  			$3^{3}S_{1}$  & $1^{--}$  & 10.354  & 10.355  & 10.356 & 10.357 & 10.343  & 10.373  & 10.349  & 10.354  & 10.334  & 10.351\tabularnewline
  			$4^{1}S_{0}$  & $0^{-+}$  & 10.638  &  & 10.597 & 10.627 & 10.584  &  &  & 10.623  &  & 10.595 \tabularnewline
  			$4^{3}S_{1}$  & $1^{--}$  & 10.650  & 10.579  & 10.612 & 10.637 & 10.597  & 10.654  & 10.607  & 10.635  &  & 10.611 \tabularnewline
  			$5^{1}S_{0}$  & $0^{-+}$  & 10.901  &  & 10.810 & 10.878 & 10.800  &  &  & 10.869  &  & 10.817 \tabularnewline
  			$5^{3}S_{1}$  & $1^{--}$  & 10.912  & 10.885  & 10.822 & 10.887 & 10.811  & 10.999  & 10.818  & 10.878  &  & 10.831 \tabularnewline
  			$6^{1}S_{0}$  & $0^{-+}$  & 11.140  &  & 10.991 & 11.111 & 10.988  &  &  & 11.097  &  & 11.011 \tabularnewline
  			$6^{3}S_{1}$  & $1^{--}$  & 11.151  & 11.000$^{\star}$ & 11.001  & 11.118 & 10.997  & 11.265  & 10.995  & 11.102  &  & 11.023 \tabularnewline
  			\midrule
  			\addlinespace
  			$1^{3}P_{0}$  & $0^{++}$  & 9.874  & 9.859  & 9.858 & 9.854 & 9.864  & 9.851  & 9.855  & 9.847  & 9.865  & 9.865 \tabularnewline
  			$1^{3}P_{1}$  & $1^{++}$  & 9.894  & 9.893  & 9.889 & 9.893 & 9.903  & 9.890  & 9.874  & 9.876  & 9.893  & 9.897 \tabularnewline
  			$1^{1}P_{1}$  & $1^{+-}$  & 9.899  & 9.899  & 9.894 & 9.899 & 9.909  & 9.886  & 9.879  & 9.882  & 9.900  & 9.903 \tabularnewline
  			$1^{3}P_{2}$  & $2^{++}$  & 9.907  & 9.912  & 9.910 & 9.911 & 9.921  & 9.899  & 9.886  & 9.897  & 9.913  & 9.918 \tabularnewline
  			$2^{3}P_{0}$  & $0^{++}$  & 10.248  & 10.233  & 10.235 & 10.239 & 10.220  & 10.233  & 10.221  & 10.226  & 10.194  & 10.226 \tabularnewline
  			$2^{3}P_{1}$  & $1^{++}$  & 10.265  & 10.255  & 10.255 & 10.259 & 10.249  & 10.257  & 10.236  & 10.246  & 10.212  & 10.251\tabularnewline
  			$2^{1}P_{1}$  & $1^{+-}$  & 10.268  & 10.260  & 10.259 & 10.262 & 10.254  & 10.262  & 10.240  & 10.250  & 10.219  & 10.256 \tabularnewline
  			$2^{3}P_{2}$  & $2^{++}$  & 10.274  & 10.269  & 10.269 & 10.268 & 10.264  & 10.274  & 10.246  & 10.261  & 10.227  & 10.269 \tabularnewline
  			$3^{3}P_{0}$  & $0^{++}$  & 10.551  &  & 10.513 & 10.551 & 10.490  & 10.533  & 10.500  & 10.522  &  & 10.502 \tabularnewline
  			$3^{3}P_{1}$  & $1^{++}$  & 10.567  & 10.512  & 10.527 & 10.557 & 10.515  & 10.556  & 10.513  & 10.538  &  & 10.524 \tabularnewline
  			$3^{1}P_{1}$  & $1^{+-}$  & 10.570  &  & 10.530 & 10.556 & 10.519  & 10.560  & 10.516  & 10.541  &  & 10.529\tabularnewline
  			$3^{3}P_{2}$  & $2^{++}$  & 10.576  & 10.524  & 10.539 & 10.556 & 10.528  & 10.568  & 10.521  & 10.550  &  & 10.540 \tabularnewline
  			\midrule
  			\addlinespace
  			$1^{3}D_{1}$  & $1^{--}$  & 10.145  &  & 10.153 & 10.136 & 10.146  & 10.136  & 10.117  & 10.138  & 10.150  & 10.145 \tabularnewline
  			$1^{3}D_{2}$  & $2^{--}$  & 10.149  & 10.163  & 10.162 & 10.164 & 10.153  & 10.141  & 10.122  & 10.147  & 10.161  & 10.151\tabularnewline
  			$1^{1}D_{2}$  & $2^{-+}$  & 10.149  &  & 10.163 & 10.167 & 10.153  &  & 10.123  & 10.148  & 10.163  & 10.152 \tabularnewline
  			$1^{3}D_{3}$  & $3^{--}$  & 10.150  &  & 10.170 & 10.183 & 10.157  &  & 10.127  & 10.155  & 10.172  & 10.156\tabularnewline
  			$2^{3}D_{1}$  & $1^{--}$  & 10.462  &  & 10.442 & 10.467 & 10.425  & 10.454  & 10.414  & 10.441  & 10.458  & 10.432 \tabularnewline
  			$2^{3}D_{2}$  & $2^{--}$  & 10.465  &  & 10.450 & 10.476 & 10.432  &  & 10.418  & 10.449  & 10.401  & 10.438 \tabularnewline
  			$2^{1}D_{2}$  & $2^{-+}$  & 10.465  &  & 10.450 & 10.475 & 10.432  &  & 10.419  & 10.450  & 10.447  & 10.439 \tabularnewline
  			$2^{3}D_{3}$  & $3^{--}$  & 10.466  &  & 10.456 & 10.478 & 10.436  &  & 10.422  & 10.455  & 10.459  & 10.442 \tabularnewline
  			$3^{3}D_{1}$  & $1^{--}$  & 10.736  & 10.752  & 10.675 & 10.742 &  & 10.725  & 10.653  & 10.698  &  & 10.670\tabularnewline
  			$3^{3}D_{2}$  & $2^{--}$  & 10.740  &  & 10.681 & 10.744 &  &  & 10.657  & 10.705  &  & 10.676\tabularnewline
  			$3^{1}D_{2}$  & $2^{-+}$  & 10.740  &  & 10.681 & 10.742 &  &  & 10.658  & 10.706  &  & 10.677\tabularnewline
  			$3^{3}D_{3}$  & $3^{--}$  & 10.741  &  & 10686 & 10.740 &  &  & 10.660  & 10.711  &  & 10.680\tabularnewline
  			$4^{3}D_{1}$  & $1^{--}$  & 10.985  &  11.000$^\dagger$ & 10.871  & 10.987 &  &  & 10.853  & 10.928 &  & 10.877 \tabularnewline
  			$4^{3}D_{2}$  & $2^{--}$  & 10.988  &  & 10.876 & 10.986 &  &  &  & 10.934  &  & 10.882 \tabularnewline
  			$4^{1}D_{2}$  & $2^{-+}$  & 10.988  &  & 10.876 & 10.984 &  &  &  & 10.935  &  & 10.883 \tabularnewline
  			$4^{3}D_{3}$  & $3^{--}$  & 10.990  &  & 10.880 & 10.981 &  &  & 10.860 & 10.939  &  & 10.886 \tabularnewline
  			\bottomrule
  	\end{tabular}}

  $\star \Upsilon\left(11020\right)$ Widely   considered. $^\dagger\Upsilon\left(11020\right)$ More favoured in the present work.
  \end{table*}
\begin{table}
	\centering
	\caption{Masses of mixed states in bottomonium  (in GeV).\label{tab:bbmixing}}
	\begin{tabular}{cccc}
		\toprule
		\multirow{2}{*}{Expt. State} & \multirow{2}{*}{Mixed State} & \multicolumn{2}{c}{Mass of mixed states (GeV)}\tabularnewline
		\cmidrule{3-4} \cmidrule{4-4}
		&  & Present & Expt\cite{Zyla:2020zbs}\tabularnewline
		\midrule
		$\Upsilon\left(10860\right)$ & $4^{3}S_{1}\text{ and }4^{3}D_{1}$ & 10.886 & $10.885_{-1.6}^{+2.6}$\tabularnewline
		
		$\Upsilon(11020)$ & $6^{3}S_{1}\text{ and }4^{3}D_{1}$ & 11.004 & $11.000\pm0.004$\tabularnewline
		
		\bottomrule
	\end{tabular}
\end{table}

    \begin{table}[h]
     \begin{center}
    \caption{Pseudoscalar and vector decay constants of the bottomonium (in $\mathrm{MeV}$)\label{tab:decaybb}}
    \begin{tabular}{cccccccc}
    \hline
    \addlinespace[2pt]
    Decay& State & Our Work &Expt.Ref.~\cite{Zyla:2020zbs}&Ref.~\cite{Pandya:2021}&Ref.~\cite{Soni:2017}& Ref.~\cite{Bhaghyesh:2011} & Ref.~\cite{Wang:2006}    \tabularnewline
    \addlinespace[2pt]
    \hline
    \addlinespace[3pt]
   \(f_V\)&$1^3S_1$ & 640(530) &715$\pm$5&653(551)&647&  931(645)  & 498$\pm$20 \tabularnewline
   &$2^3S_1$ & 383(317) &498$\pm$8 &563(477)&519& 566(439) &366$\pm$27 \tabularnewline
   &$3^3S_1$ & 338(280) &430$\pm$4&507(430)&475& 507(393) & 304$\pm$27 \tabularnewline
   &$4^3S_1$ & 319(265) &336$\pm$18&466(395)&450&481(356)  &259$\pm$22 \tabularnewline
  &$5^3S_1$ & 308(255) &&434(368)&432& 439(341) & 228$\pm$16\tabularnewline
   &$6^3S_1$ & 301(249) &&406(345)&418&  & \tabularnewline
     \addlinespace[2pt]
     \hline
     \addlinespace[3pt]
    \(f_P\) & $1^1S_0$ & 639(529) &&654(578)&646&834(694)   &   \tabularnewline
    & $2^1S_0$ & 382(317) &&564(499)&518&567(472)  & \tabularnewline
    & $3^1S_0$ & 338(280) &&508(450)&474& 508(422) &  \tabularnewline
    & $4^1S_0$ & 319(264) &&467(413)&449& 481(401) & \tabularnewline
    & $5^1S_0$ & 308(255) &&435(385)&432&  & \tabularnewline
    & $6^1S_0$ & 301(249) &&406(360)&418&  & \tabularnewline

    \hline
    \end{tabular}
    \end{center}
    \end{table}

      \begin{table*}
      \centering{}\caption{Electric dipole (E1) transitions widths of $b\overline{b}$ meson. \label{tab:E1BB}}
       \begin{tabular}{ccccccccccccc}
      \hline
      \addlinespace[2pt]
      \multicolumn{2}{c}{Meson trans.} & \multicolumn{2}{c}{This work} & Expt. Ref.\cite{Zyla:2020zbs} & \multicolumn{8}{c}{Other work ($\varGamma$ in keV)}\tabularnewline
      \addlinespace[2pt]
      \cline{3-4}\cline{6-13}
      \addlinespace[3pt]
      Initial & Final & $E_{\gamma}$ & $\varGamma$ & $\varGamma$ &Ref.&Ref.& Ref.  & Ref.& Ref. & Ref. &Ref. &Ref.\tabularnewline
       & & MeV& keV & keV & \cite{Wang:2018}&\cite{Segovia2016}& \cite{Ebert2003}  & \cite{Akbar:2015}& \cite{Godfrey:2015} & \cite{Chaturvedi:2020} & \cite{Brambilla:2004}&\cite{Pandya:2021} \tabularnewline
      \addlinespace[2pt]
       \hline
       \addlinespace[5pt]
      $1^{3}P_{2}$ & $1^{3}S_{1}$ & 434 & 39.99 & 34.38 &31.4&39.15& 40.2 & 37.3 &  32.8 & 39.27 & 31.6&15.70\tabularnewline
      $1^{3}P_{1}$ & $1^{3}S_{1}$ & 422 & 36.80 & 32.544 &28.3&35.66& 36.6 & 32.8 &   29.5 & 42.55 & 27.8&14.66\tabularnewline
      $1^{1}P_{1}$ & $1^{1}S_{0}$ & 465 & 24.68 & 35.77 &34.4&43.66& 52.6 & 22.9 &   35.7 & 74.63 & 41.8&12.52\tabularnewline
      $1^{3}P_{0}$ & $1^{3}S_{1}$ & 403 & 16.00 &  &22.8&28.07& 29.9 & 26.0 &  23.8 & 29.07 & 22.2&19.4\tabularnewline
       \addlinespace[2pt]
        \hline
        \addlinespace[5pt]
      $2^{3}S_{1}$ & $1^{3}P_{2}$ & 105 & 1.78 & $2.29\pm0.23$ &1.86&2.08& 2.46 & 2.92 &   1.88 &2.274  & 2.04&0.841\tabularnewline
      $2^{3}S_{1}$ & $1^{3}P_{1}$ & 119 & 1.49 & $2.21\pm0.22$ &1.60&1.84& 2.45 & 2.83 &   1.63 & 2.232 & 2.00&0.614\tabularnewline
      $2^{3}S_{1}$ & $1^{3}P_{0}$ & 124 & 0.86 & $1.22\pm0.16$ &0.907&1.09& 1.62 & 1.85 &  0.91 & 1.30 & 1.29&0.302\tabularnewline
      $2^{1}S_{0}$ & $1^{1}P_{1}$ & 84 & 1.60 &  &2.467&2.85& 3.09 & 4.54 &   2.48 & 8.045 & 41.8&0.888\tabularnewline
       \addlinespace[2pt]
        \hline
        \addlinespace[5pt]
      $1^{3}D_{3}$ & $1^{3}P_{2}$ & 240 & 21.68 &  &23.9&24.74& 24.6 & 31.27 &   24.3 & 5.099 & 22.6&6.677\tabularnewline
      $1^{3}D_{2}$ & $1^{3}P_{2}$ & 239 & 5.34 &  &5.49&6.23& 6.35 & 5.48 &   5.6 & 3.017 & 5.46&1.57\tabularnewline
      $1^{3}D_{2}$ & $1^{3}P_{1}$ & 251 & 20.63 &  &18.8&21.95& 23.3 & 21.03 &  19.2 & 3.940 & 20.5&5.242\tabularnewline
      $1^{3}D_{1}$ & $1^{3}P_{2}$ & 236 & 0.57 &  &0.55&0.65& 0.69 & 0.568   & 0.56 & 1.908 & 0.50&0.161\tabularnewline
      $1^{3}D_{1}$ & $1^{3}P_{1}$ & 248 & 9.91 &  &9.51&12.29& 12.7 & 10.96   & 9.7 & 2.264 & 10.7&2.684\tabularnewline
      $1^{3}D_{1}$ & $1^{3}P_{0}$ & 267 & 16.58 &  &16.3& 20.98& 23.4 & 25.84 &   16.5 & 3.120 & 20.1&4.476\tabularnewline
       \addlinespace[2pt]
        \hline
        \addlinespace[5pt]
      $2^{3}P_{2}$ & $2^{3}S_{1}$ & 272 & 20.02 & 24.645 &14.6&17.50& 16.7 & 18.99 &   14.3 &15.66 & 14.5 &\tabularnewline
      $2^{3}P_{1}$ & $2^{3}S_{1}$ & 262 & 18.03 & 23.283 &13.7&15.89& 14.7 & 16.14 &   13.3 & 21.84 & 12.4 & \tabularnewline
      $2^{1}P_{1}$ & $2^{1}S_{0}$ & 281 & 22.09 & 40.32 &15.0&17.6& 21.4 & 12.79 &   14.1 & 23.11 & 19& \tabularnewline
      $2^{3}P_{0}$ & $2^{3}S_{1}$ & 247 & 14.96 & 0.00012 &11.1&12.8& 6.79 & 11.88 &   10.9 & 16.84 & 9.17&\tabularnewline
       \addlinespace[2pt]
        \hline
        \addlinespace[5pt]
      $2^{3}P_{2}$ & $1^{3}D_{3}$ & 124 & 3.097 &  &1.61&2.06& 2.35 & 2.885 &   1.5 &  & 2.25& \tabularnewline
      $2^{3}P_{2}$ & $1^{3}D_{2}$ & 125 & 0.569 &  &0.339&0.35& 0.449 & 1.054 &   0.3 &  & 0.434& \tabularnewline
      $2^{3}P_{2}$ & $1^{3}D_{1}$ & 128 & 0.041 &  &0.027&0.021& 0.035 & 0.079 &   0.03 &  & 0.034& \tabularnewline
      $2^{3}P_{1}$ & $1^{3}D_{1}$ & 119 & 0.818 &  &0.511&0.41& 0.615 & 1.46 &   0.5 &  & 0.593& \tabularnewline
      $2^{3}P_{0}$ & $1^{3}D_{1}$ & 103 & 2.11 &  &1.05&0.74& 1.17 & 3.21 &   1.0 &  & 1.13&\tabularnewline
       \addlinespace[2pt]
        \hline
       \end{tabular}
      \end{table*}

     \begin{table*}
     \begin{centering}
     \caption{Magnetic dipole (M1) transitions widths of $b\overline{b}$ mesons.\label{tab:M1BB}} \smallskip
     \par\end{centering}
     \centering{}%
     \begin{tabular}{ccccccccc}
     \hline
     \addlinespace[3pt]
     Meson transition & \multicolumn{2}{c}{This work} & \multicolumn{6}{c}{Other work ($\varGamma$ in keV)}\tabularnewline
      \addlinespace[2pt]
     \cline{2-3}
      \addlinespace[2pt]
     Initial$\rightarrow$Final & $E_{\gamma}$(MeV) & $\varGamma$(keV) & Ref.~\cite{Wang:2018}& Ref.~\cite{Segovia2016} &Ref.~\cite{Akbar:2015} & Ref.~\cite{Godfrey:2015}   & Ref.~\cite{Brambilla:2004} & Ref.~\cite{Ebert:2011jc}\tabularnewline
    \addlinespace[2pt]
       \hline
      \addlinespace[4pt]
     $1^{3}S_{1}\rightarrow1^{1}S_{0}$ & 41 & $9.1\times10^{-3}$ &$9.52\times10^{-3}$ &$9.34\times10^{-3}$&$11\times10^{-3}$  & $10\times10^{-3}$  & $8.95\times10^{-3}$ &$9.7\times10^{-3}$ \tabularnewline

     $2^{3}S_{1}\rightarrow2^{1}S_{0}$ & 15 & $4.98\times10^{-4}$ &$5.82\times10^{-4}$ &$5.80\times10^{-4}$&$6.6\times10^{-4}$   & $5.9\times10^{-4}$ & $15.1\times10^{-4}$ & $1.6\times10^{-3}$ \tabularnewline

     $3^{3}S_{1}\rightarrow3^{1}S_{0}$ & 12 & $4.4\times10^{-4}$ &$3.37\times10^{-4}$ &$6.58\times10^{-4}$&0.012   & $2.5\times10^{-4}$  & $8.3\times10^{-4}$ & $0.9\times10^{-3}$ \tabularnewline

     $2^{3}S_{1}\rightarrow1^{1}S_{0}$ & 560 & 1.609 & 0.0688&0.0565&0.1729   & 0.081   & 0.00281 & $1.3\times10^{-3}$ \tabularnewline

     $2^{1}S_{0}\rightarrow1^{3}S_{1}$ & 506 & 3.585 &0.0706 &0.045&0.00064   & 0.068   & 0.00283 & $2.4\times10^{-3}$\tabularnewline
      \addlinespace[2pt]
     \hline
     \end{tabular}
     \end{table*}

    \begin{table*}
    \caption{Leptonic decay widths ({$\Upsilon \rightarrow\varGamma_{l^{+}l^{-}}$} in keV and {$n^{1}D_{2} \rightarrow\varGamma_{l^{+}l^{-}}$} in eV ).\label{annihi2e}}
       \centering{}%
       \begin{tabular}{cccccccccccccc}
      \hline
      \addlinespace[2pt]
      \multirow{3}{*}{State} & \multicolumn{2}{c}{This work} &  Expt.& \multicolumn{10}{c}{Other works } \tabularnewline
       \addlinespace[2pt]
      \cline{2-3} \cline{5-14}
       \addlinespace[2pt]

      & \multirow{2}{*}{$\varGamma_{l^{+}l^{-}}$} & \multirow{2}{*}{$\varGamma_{l^{+}l^{-}}^{cf}$}& Ref.& Ref. & Ref. & Ref.& Ref. & Ref. & Ref. & Ref. & Ref. & Ref. & Ref.\tabularnewline
      &  &   & \cite{Zyla:2020zbs}& \cite{Wang:2018} & \cite{Segovia2016} & \cite{Godfrey:2015}& \cite{Bhaghyesh:2011}&\cite{Chaturvedi:2020} & \cite{Giannuzzi:2008} & \cite{Radford:2007} & \cite{Li:2009nr} &\cite{Gonzalez:2004} & \cite{Pandya:2021} \tabularnewline
       \addlinespace[2pt]
     \hline
        \addlinespace[3pt]
     $\Upsilon(1^{3}S_1)$ & 1.074 & 0.582 & $1.340\pm0.018$ &1.65&  0.71&1.44 & 1.809&1.053 & 1.237 & 1.33 & 2.31 & 1.01 & 1.224 \tabularnewline
     $\Upsilon(2^{3}S_1)$ & 0.363 & 0.197 & $0.612\pm0.011$ &0.821& 0.37& 0.73& 0.797&0.562 & 0.581 & 0.61 & 0.92 & 0.35 & 0.537 \tabularnewline
     $\Upsilon(3^{3}S_1)$ & 0.274 & 0.149 & $0.443\pm0.008$& 0.569& 0.27&0.53 & 0.618 &0.399& 0.270 & 0.46 & 0.64 & 0.25 & 0.402 \tabularnewline
     $\Upsilon(4^{3}S_1)$ & 0.238 & 0.129 & $0.272\pm0.029$ &0.431& 0.21&0.39 & 0.541 &0.282& 0.212 & 0.35 & 0.51 & 0.22 & 0.240\tabularnewline
     $\Upsilon(5^{3}S_1)$ & 0.216 & 0.117 & $0.31\pm0.07$ &0.348& 0.18 &0.33& 0.481 &0.212&  &  & 0.42 & 0.18 &  \tabularnewline
     $\Upsilon(6^{3}S_1)$ & 0.201 & 0.109 & $0.130\pm0.03$ &0.286& 0.15&0.27 & 0.432 &0.166&  &  & 0.37 & 0.15 &  \tabularnewline
       \addlinespace[2pt]
       \hline
        \addlinespace[3pt]
     $1^{3}D_{1}$ & 3.04  & 1.65  &  &1.88& 1.40&1.38  & & &  &  &  &  &  \tabularnewline
     $2^{3}D_{1}$ & 4.47  & 2.42  &  &2.81& 2.50 & 1.99& & &  &  &  &  &  \tabularnewline
     $3^{3}D_{1}$ & 5.88  & 3.19  &  & 3.0 && 2.38 &  &  &&  & & &   \tabularnewline
     $4^{3}D_{1}$ & 7.33  & 3.97  &  & 3.0 && 2.18 &  &  & & & & &   \tabularnewline
      \addlinespace[2pt]
       \hline
      \end{tabular}
      \end{table*}

  \begin{table*}
     \caption{Di-photon decay widths .\label{annihi2p}}
     \centering{}%
     \scalebox{0.88}{
     \begin{tabular}{cccccccccccccc}
      \addlinespace[2pt]
      \hline
      \addlinespace[3pt]
      \multirow{3}{*}{State} & \multicolumn{2}{c}{This work} &  \multicolumn{10}{c}{Other works ($\varGamma$ in keV)} \tabularnewline
     \addlinespace[2pt]
     \cline{2-3} \cline{5-14}
    \addlinespace[2pt]
      & $\varGamma_{\gamma\gamma}$ & $\varGamma_{\gamma\gamma}^{cf}$ &&Ref.& Ref.& Ref. & Ref. & Ref. & Ref. & Ref. & Ref. & Ref. & Ref.  \tabularnewline
      &(keV) & (keV) &&\cite{Wang:2018}& \cite{Segovia2016}& \cite{Godfrey:2015} & \cite{Anisovich:2005} & \cite{Munz:1996} & \cite{Ebert:2003b} & \cite{Pandya:2021} & \cite{Li:2009nr} & \cite{Laverty:2009} & \cite{Chaturvedi:2020} \tabularnewline
      \addlinespace[2pt]
      \hline
      \addlinespace[3pt]
     $\eta_{b}(1^{1}S_0)$ & 0.3364 & 0.2361 &&1.05& 0.69 &0.94 &1.554 & 0.22 & 0.35 & 0.378 & 0.527 & 0.30 & 0.545  \tabularnewline
     $\eta_{b}(2^{1}S_0)$ & 0.1270 & 0.0896 &&0.489& 0.36 &0.41 &1.928 & 0.110 & 0.15 & 0.263 & 0.263 & 0.14 & 0.124  \tabularnewline
     $\eta_{b}(3^{1}S_0)$ & 0.1029 & 0.0726 &&0.323& 0.27 &0.29& 2.139 & 0.084 & 0.10 & 0.206 & 0.172 & 0.10 & 0.105 \tabularnewline
     $\eta_{b}(4^{1}S_0)$ & 0.0942 & 0.0666 && 0.237& & 0.20 &&  &  & 0.170 & 0.105 &  & 0.068  \tabularnewline
     $\eta_{b}(5^{1}S_0)$ & 0.0900 & 0.0636 &&0.192 & & 0.17 &  &  && 0.145 & 0.121 &  & 0.060  \tabularnewline
     $\eta_{b}(6^{1}S_0)$ &  0.0876 &  0.0619& & 0.152& & 0.41 &  &&  & 0.123 & 0.050 &  & 0.048  \tabularnewline
      \addlinespace[2pt]
      \hline
      \addlinespace[3pt]
     $1^{3}P_{0}$ & 0.0165 & 0.0168& &0.199 &0.12 &0.15& 0.024 & 0.024 & 0.038 & 0.081  & 0.037 & 0.0329 & 0.068\tabularnewline
     $2^{3}P_{0}$ &  0.0169 & 0.0172& &0205 &0.14 &0.15& 0.023 & 0.026 & 0.029 & 0.071 & 0.037 & 0.0314 & 0.022\tabularnewline
     $3^{3}P_{0}$ & 0.0189 & 0.0192 &&0.180& 0.15 &0.13& 0.023 &  &  & 0.062 & 0.035 &  &0.008\tabularnewline
      \addlinespace[2pt]
      \hline
      \addlinespace[3pt]
     $1^{3}P_{2}$ & 0.00439 & 0.00238 &&0.0106& 0.00308 &0.0093& 0.016 & 0.0056 & 0.008 & 0.021 & 0.0066 & 0.00719 & 0.014\tabularnewline
     $2^{3}P_{2}$ & 0.00451 & 0.00245& &0.0133& 0.00384 &0.012& 0.015 & 0.0068 & 0.006 & 0.019 & 0.0067 & 0.00759 & 0.004\tabularnewline
     $3^{3}P_{2}$ & 0.00504 & 0.00273 & &0.0141& 0.0041 &0.013& 0.015 &  &  & 0.016 & 0.0064 &  &0.002 \tabularnewline
      \addlinespace[2pt]
      \hline
     \end{tabular}}
     \end{table*}

          \begin{table}
          \caption{Tri-photon decay widths (in eV).\label{annihi3p}}
           \centering{}%
          \begin{tabular}{cccccccc}
          \addlinespace[2pt]
          \hline
          \addlinespace[3pt]
          \multirow{2}{*}{State}  & \multicolumn{2}{c}{This work} &  \multicolumn{4}{c}{Other works } \\
           \addlinespace[2pt]
            \cline{2-3} \cline{5-8}
            \addlinespace[3pt]
           & $\varGamma_{3\gamma}$ & $\varGamma_{3\gamma}^{cf}$& & Ref.~\cite{Wang:2018}& Ref.~\cite{Segovia2016}& Ref.~\cite{Godfrey:2015}&Ref.~\cite{Chaturvedi:2020}\tabularnewline
          \addlinespace[2pt]
          \hline
          \addlinespace[3pt]
         $\Upsilon(1^{3}S_1)$ & $33.56\times10^{-3}$ & $30.67\times10^{-3}$ &&$19.4\times10^{-3}$& $3.44\times10^{-3}$ & $17.0\times10^{-3}$&$16\times10^{-3}$\tabularnewline
         $\Upsilon(2^{3}S_1)$ & $12.67\times10^{-3}$ & $11.58\times10^{-3}$& &$10.9\times10^{-3}$& $2.00\times10^{-3}$& $9.8\times10^{-3}$&$3\times10^{-3}$\tabularnewline
         $\Upsilon(3^{3}S_1)$ & $10.261\times10^{-3}$ & $9.376\times10^{-3}$ &&$8.04\times10^{-3}$& $1.55\times10^{-3}$& $7.6\times10^{-3}$&$1\times10^{-3}$\tabularnewline
         $\Upsilon(4^{3}S_1)$ &  $9.400\times10^{-3}$ & $8.590\times10^{-3}$ &&$6.36\times10^{-3}$ & $1.29\times10^{-3}$& $6.0\times10^{-3}$& \tabularnewline
         $\Upsilon(5^{3}S_1)$ & $8.979\times10^{-3}$ & $8.206\times10^{-3}$ &&$5.43\times10^{-3}$& $1.10\times10^{-3}$& &\tabularnewline
         $\Upsilon(6^{3}S_1)$ & $8.735\times10^{-3}$ & $7.982\times10^{-3}$ &&$4.57\times10^{-3}$& $9.56\times10^{-4}$&& \tabularnewline
           \addlinespace[2pt]
           \hline
          \end{tabular}
          \end{table}

      \begin{table*}
      \caption{Di-gluon decay widths.\label{annihi2g}}
       \centering{}%

      \begin{tabular}{cccccccccc}
      \addlinespace[2pt]
      \hline
      \addlinespace[3pt]
      \multirow{3}{*}{State}  & \multicolumn{2}{c}{This work} &   \multicolumn{7}{c}{Other works } \\
        \addlinespace[2pt]
          \cline{4-10 }
       \addlinespace[2pt]
      & $\varGamma_{gg}$ & $\varGamma_{gg}^{cf}$ &Ref.& Ref. & Ref. & Ref. & Ref. & Ref. & Ref. \tabularnewline
       & (MeV) & (MeV) &\cite{Wang:2018}& \cite{Segovia2016} & \cite{Godfrey:2015} & \cite{Pandya:2021}  & \cite{Laverty:2009} & \cite{Ebert2003,Ebert:2011jc} &\cite{Negash:2015} \tabularnewline
      \addlinespace[2pt]
      \hline
      \addlinespace[3pt]
      $\eta_{b}(1^{1}S_0)$ & 8.219 & 11.326 &17.9& 20.18 &16.6& 5.449     & 11.49 &  & 10.865\tabularnewline
      $\eta_{b}(2^{1}S_0)$ & 3.121 & 4.301 & 8.33&10.64& 7.2& 4.177&     5.16 &  &7.477 \tabularnewline
      $\eta_{b}(3^{1}S_0)$ & 2.529 & 3.485 & 5.51&7.94& 4.9&3.449 &   3.81 &  &6.276 \tabularnewline
      $\eta_{b}(4^{1}S_0)$ & 2.317 & 3.193 & 4.03 &  &3.4&2.945&    &  & \tabularnewline
      $\eta_{b}(5^{1}S_0)$ & 2.214 & 3.051 & 3.26 &  &2.9&2.576  &  &&   \tabularnewline
      $\eta_{b}(6^{1}S_0)$ &  2.154 & 2.968 & 2.59 & &2.2&2.285   &  &  & \tabularnewline
       \addlinespace[2pt]
       \hline
       \addlinespace[3pt]

      $1^{3}P_{0}$ & 0.721  & 1.34 &3.37 &2.00& 2.6& 0.906   & 0.96 & 0.653  &   \tabularnewline
      $2^{3}P_{0}$ & 0.741  & 1.39 &3.52 &2.37&2.29 &0.785  & 0.99 & 0.431  &   \tabularnewline
      $3^{3}P_{0}$ & 0.828 & 1.54 &3.10 &2.46& 2.2& 0.671   &  &  & \tabularnewline
       \addlinespace[2pt]
       \hline
       \addlinespace[3pt]
      $1^{3}P_{2}$ & 0.192 & 0.209 & 0.165&0.08369& 0.147& 0.238 & 0.33 & 0.109 &   \tabularnewline
      $2^{3}P_{2}$ & 0.198 & 0.215 &0.220&0.10426&0.207 & 0.207 & 0.35 & 0.076  &  \tabularnewline
      $3^{3}P_{2}$ & 0.221 & 0.240 & 0.243&0.11145&0.227 & 0.178   &   & & \tabularnewline
       \addlinespace[2pt]
       \hline
       \addlinespace[3pt]
      $1^{1}D_{2}$ & 0.489 (keV) &  & 0.657(keV)&0.37(keV)& 1.87(keV)&    &  &  & \tabularnewline
      $2^{1}D_{2}$ & 0.764 (keV) &  &1.22(keV) &0.67(keV)&3.3(keV) &    &  &  & \tabularnewline
      $3^{1}D_{2}$ & 1.06 (keV) &  &  1.59(keV)&  & 4.7(keV) &&  &  & \tabularnewline
      $4^{1}D_{2}$ & 1.38 (keV) &  & 1.86(keV) &  &  &&&  &   \tabularnewline
      \addlinespace[2pt]
      \hline
      \end{tabular}
      \end{table*}

      \begin{table}
       \caption{Tri-gluon decay widths (in keV).\label{annihi3g}}
       \centering{}%
       \begin{tabular}{cccccccc}
         \hline
         \addlinespace[3pt]
         \multirow{2}{*}{State} & \multicolumn{2}{c}{This work} &   \multicolumn{4}{c}{Other works} \tabularnewline
          \addlinespace[2pt]
           \cline{4-8}
          \addlinespace[3pt]
        & $\varGamma_{3g}$ & $\varGamma_{3g}^{cf}$ &Ref.~\cite{Wang:2018}& Ref.~\cite{Segovia2016}&Ref.~\cite{Godfrey:2015}&Ref.~\cite{Pandya:2021}  & Ref.~\cite{Ebert:2011jc} \tabularnewline
         \addlinespace[2pt]
         \hline
         \addlinespace[3pt]
       $\Upsilon(1^{3}S_1)$ & 114.73  & 66.42  &50.8& 41.63 &47.6&  40.0& \tabularnewline
       $\Upsilon(2^{3}S_1)$ & 43.33 & 25.08 &28.4& 24.25 &26.3& 26.9&  \tabularnewline
       $\Upsilon(3^{3}S_1)$ & 35.08  & 20.31  &21.0 &18.76 &19.8& 20.6&  \tabularnewline
       $\Upsilon(4^{3}S_1)$ & 32.14 & 18.65  &16.7& 15.58 &15.1& 16.8& \tabularnewline
       $\Upsilon(5^{3}S_1)$ & 30.70  & 17.78  &14.2& 13.33 &13.1& 14.1& \tabularnewline
       $\Upsilon(6^{3}S_1)$ & 29.87 & 17.29 &12.0& 11.57 &11.0&  11.7& \tabularnewline
         \addlinespace[2pt]
         \hline
         \addlinespace[3pt]
       $1^{1}P_{1}$ & 28.47 &  &44.7& 35.26 &37.0 & 35.7&36 \tabularnewline
       $2^{1}P_{1}$ & 35.10 &  & 64.6&52.70 & 54.0& 34.6&31.5\tabularnewline
       $3^{1}P_{1}$ & 43.11 &  & 71.1&62.16 &59.0 &33.1 &\tabularnewline
         \addlinespace[2pt]
         \hline
         \addlinespace[3pt]
       $1^{3}D_{1}$ & 9.64  &  &10.4& 9.97 &8.11& 10.6 &\tabularnewline
       $2^{3}D_{1}$ & 16.27 &  & 20.1&9.69 &14.8 &11.9 & \tabularnewline
       $3^{3}D_{1}$ & 23.67  &  &26.0 & &  21.2&11.8 &\tabularnewline
       $4^{3}D_{1}$ & 31.99  &  &30.4 & &  &11.3 &\tabularnewline
         \addlinespace[2pt]
         \hline
         \addlinespace[3pt]
       $1^{3}D_{2}$ & 0.482 &  & 0.821&0.62 &0.69 && \tabularnewline
       $2^{3}D_{2}$ & 0.813 &  &1.65 &0.61 &1.4 &  &\tabularnewline
       $3^{3}D_{2}$ & 1.18 &  & 2.27 && 2.0 && \tabularnewline
       $4^{3}D_{2}$ & 1.60 &  & 2.75 && &  &\tabularnewline
         \addlinespace[2pt]
         \hline
         \addlinespace[3pt]
       $1^{3}D_{3}$ & 1.927  &  &2.19 &0.22 & 2.07&6.0 &\tabularnewline
       $2^{3}D_{3}$ & 3.252  &  & 4.56&1.25 &4.3 &5.6& \tabularnewline
       $3^{3}D_{3}$ & 4.732  &  & 6.65& &  6.6&5.5& \tabularnewline
       $4^{3}D_{3}$ & 6.393  &  & 8.38 && & 5.3& \tabularnewline
         \addlinespace[2pt]
         \hline
        \end{tabular}
       \end{table}

        \begin{table}
        \caption{$n^{3}S_{1}\rightarrow\gamma gg$ decay widths (in keV).\label{annihip2g}}
        \centering{}%
        \begin{tabular}{cccccccc}
        \hline
        \addlinespace[2pt]
         \multirow{2}{*}{State} & \multicolumn{2}{c}{This work} &Expt.&   \multicolumn{4}{c}{Other works } \\
         \cline{2-3} \cline{5-8}
         \addlinespace[3pt]
         & $\varGamma_{\gamma gg}$ & $\varGamma_{\gamma gg}^{cf}$&Ref.~\cite{Zyla:2020zbs} &Ref.~\cite{Wang:2018}& Ref.~\cite{Segovia2016}&Ref.~\cite{Godfrey:2015}&Ref.~\cite{Pandya:2021}\tabularnewline
       \addlinespace[2pt]
        \hline
        \addlinespace[3pt]
        $\Upsilon(1^{3}S_1)$ & 2.481 & 0.903 &1.18& 1.32&0.79&1.2&1.273\tabularnewline
        $\Upsilon(2^{3}S_1)$ & 0.937  & 0.341 &0.59&0.739& 0.46 &0.68&0.869 \tabularnewline
        $\Upsilon(3^{3}S_1)$ & 0.759 & 0.276 & 0.0097& 0.547&0.36&0.52&0.672 \tabularnewline
        $\Upsilon(4^{3}S_1)$ & 0.695 & 0.253 & &0.433 &0.30 &0.40&0.548 \tabularnewline
        $\Upsilon(5^{3}S_1)$ & 0.664 & 0.242 &&0.370 &0.25&&0.465\tabularnewline
        $\Upsilon(6^{3}S_1)$ & 0.646 & 0.235 &&0.311& 0.22&&0.389 \tabularnewline
       \addlinespace[2pt]
        \hline
        \end{tabular}
        \end{table}

        \begin{table}
        \caption{$n^{3}P_{1}\rightarrow q\overline{q}+g$ decay widths (in keV).\label{annihiqqg}}
         \centering{}%
        \begin{tabular}{ccccccc}
        \hline
        \addlinespace[3pt]
         \multirow{2}{*}{State} & This work &  \multicolumn{5}{c}{Other works } \\

          \cline{3-7}
         \addlinespace[3pt]
         & $\varGamma_{q\overline{q}+g}$ &Ref.~\cite{Wang:2018}& Ref.~\cite{Segovia2016} &Ref.~\cite{Godfrey:2015} &Ref.~\cite{Ebert2003}&Ref.~\cite{Pandya:2021}\tabularnewline
       \addlinespace[2pt]
        \hline
        \addlinespace[3pt]
        $1^{3}P_{1}$ & 45.55& 81.7& 71.53 &67& 57&57.958\tabularnewline
        $2^{3}P_{1}$ &  56.16 &117 & 106.14& 96& 50&55.396\tabularnewline
        $3^{3}P_{1}$ & 68.97 &126& 124.53 && &52.958\tabularnewline
        \addlinespace[2pt]
         \hline
        \end{tabular}
        \end{table}


\section{Results and Discussion\label{sec:Resu}}

The bottomonium mass spectrum is investigated using the Cornell potential with a Gaussian wave function and  a $\mathcal{O}(1/m)$ correction to the potential energy term and up to a ${{\cal{O}}\left({\bf p}^{10}\right)}$ expansion of the kinetic energy term  for relativistic correction of the Hamiltonian.  We have calculated Hamiltonian yield (center of weight masses) for the nS $(n\leq6)$, $ nP\; \text{and}\; nD \;(n\leq4)$ bottomonium states and are presented in Table~\ref{tab:bbmsa}.

\subsection{Masses of S States}

The masses of the low lying S states are well measured experimentally. The estimated complete mass spectrum of bottomonium is presented in Table \ref{tab:massesbb}, with the spectroscopic notation $n^{2S+1}L_{J}$ and is also graphically represented in Fig.\ref{fig:MassBB}. We have compared our results with other theoretical model predictions. In Ref. \cite{Wang:2018} Wang \textit{et. al.}   employ modified Godfrey-Isgur (GI) model, in ref. \cite{Chen:2019uzm} Chen \textit{et. al.} employ the Relativistic Flux Tube  (RFT) model, in ref. \cite{Deng:2017} Deng \textit{et. al.} use the Non-Relativistic Screened Potential Model (NR-SPM), in ref. \cite{Lu:2016mbb} Lu \textit{et. al.} use the linear plus color-Coulomb (GI-type), Segovia \textit{et. al.} employ the Non-Relativistic Constituent Quark Model (NR-CQM), Godfrey \textit{et. al.} in ref. \cite{Godfrey:2015} employ Relativised Quark Model (RQM), Wurtz \textit{et. al.} in ref. \cite{Wurtz:2015}  present LATTICE Field Theory calculations while Li \textit{et. al.} also employ a Screened Potential Model (SPM) as presented in Table \ref{tab:massesbb}.  We obtained the pseudoscalar state mass $\eta_{b} \left(1^{1}S_{0}\right)$ (9423 MeV) and vector state mass $\Upsilon \left(1^{3}S_{1}\right)$ (9463 MeV), by adding the spin hyperfine interaction to the fixed spin average mass for the ground state. The estimated masses for $2^{1}S_{0}$ ($\eta_{b}$(2S)), $2^{3}S_{1}$ ($\Upsilon(2S)$) and  $3^{3}S_{1}$ ($\Upsilon(3S)$) states are underestimated by 16 MeV, 22 Mev and 1 MeV respectively when compared with the experimentally observed values.

The $\Upsilon\left(4S\right)$ is considered to be a $4^{3}S_{1}$ state with the experimental mass $10579.4 \pm 1.2$ MeV. In the present work the calculated mass $M_{\Upsilon\left(4S\right)} = 10657$ MeV   is overestimated by 71 MeV  as compared to experimentally observed mass. The calculated mass is also overestimated by other theoretical model predictions. For example the estimated mass by RFT model in ref. \cite{Chen:2019uzm} is $M_{\Upsilon\left(4S\right)} = 10637$ MeV, the modified GI model in ref. \cite{Wang:2018} is  $M_{\Upsilon\left(4S\right)} = 10612$ MeV and NR-SPM in ref. \cite{Deng:2017} is $M_{\Upsilon\left(4S\right)} = 10597$ MeV.

The state $\Upsilon\left(10860\right)$ is generally assigned to be the $5^{3}S_{1}$ state. However other possible interpretations such as a mixing of $\Upsilon\left(5S\right)$ with a lowest P-wave hybrid \cite{Bruschini2019} is also considered. The experimentally measured mass is $M_{\Upsilon\left(10860\right)} = 10885.2^{+2.6}_{-1.6}$ MeV.  The calculated mass of $5^{3}S_{1}$ state in the present work is 10912 MeV which is overestimated.   The difference between our calculated and experimentally determined mass is 27 MeV. In Table \ref{tab:bbmixing} we present the mass of $\Upsilon\left(10860\right)$ state $M_{\Upsilon\left(10860\right)} = 10886$ MeV obtained as a mixture of $4^{3}S_{1}$ and $4^{3}D_{1}$ state. Thus we believe this state could also be a mixture of S-D states. Interestingly the relativistic flux tube (RFT) model\cite{Chen:2019uzm} and lattice QCD\cite{Bicudo:2019ymo} study  favour $5^{3}S_{1}$ association for ${{\Upsilon}{(10860)}}$.

The state $\Upsilon\left(11020\right)$ is widely considered to be the $6^{3}S_{1}$ state. Its measured experimental mass is $M_{\left(\Upsilon{11020}\right)} = 11000 \pm 4$ MeV. In the present work we obtain mass of $6^{3}S_{1}$ state as $11151$  MeV which is overestimated. Thus the present model does not favour $6^{3}S_{1}$ assignment to the $\Upsilon\left(11020\right)$ resonance. Mass of $6^{3}S_{1}$ states in many other theoretical model predictions also does not match with  $\Upsilon\left(11020\right)$ state. Thus $\Upsilon\left(11020\right)$ may not be a pure 6S resonance. In ref. \cite{Badalian2010} Badalian \textit{et. al.} have considered this state to be a mixture of S-D states. In the present work mass of $4^{3}D_{1}$ state is 10985 MeV which is slightly underestimated yet closer to the $\Upsilon\left(11020\right)$ experimental mass. Thus if we consider mixing between S-D states, then our model favours mixing between $6^{3}S_{1}$ and $4^{3}D_{1}$.  We have considered it to be a  $4^{3}D_{1}$ to draw the Regge trajectory in fig. \ref{fig:PsVmesonBB}.

The hyperfine mass splitting $\Delta m(nS)=m\left[\Upsilon\left(nS\right)\right]-m\left[\eta_{b}\left(nS\right)\right]$ between the singlet and triplet S states, reflects the spin-dependent interaction. Our predicted hyperfine mass splitting between 1S and 2S states is 40 MeV and 18 MeV respectively, which are much less when compared with the experimental and with that of other theoretical models. Our predicted hyperfine mass splitting for higher bottomonium states $nS\;(n\geq3)$  are in notable difference but consistent with predictions by other theoretical models.

\subsubsection{Masses of P states}

Until now all 1P and 2P states are confirmed by experiment\cite{Zyla:2020zbs}, calculated masses for 1P-3P states can be found in table \ref{tab:bitaBB}. The mass difference between the estimated value by our model  and experimentally observed values are  15 MeV for $1^{3}P_{0}$ ($\chi_{b0}(1P)$), 1 MeV for $1^{3}P_{1}$ ($\chi_{b1}(1P)$), matches well for $1^{1}P_{1}$ ($h_{b}(1P)$), shows 5 MeV difference for $1^{3}P_{2}$ ($\chi_{b2}(1P)$),  15 MeV for $2^{3}P_{0}$ ($\chi_{b0}(2P)$), 10 MeV for $2^{3}P_{1}$ ($\chi_{b1}(2P)$), 8 MeV for $2^{1}P_{1}$ ($h_{b}(2P)$), 5 MeV for $2^{3}P_{2}$ ($\chi_{b2}(2P)$).

Two  3P states ${ \chi_{{b1}}{(3P)}}$ and ${{\chi}_{{b2}}{(3P)}}$ have been observed by ATLAS \cite{Aad:2011}, D0 \cite{Abazov:2012gh} and LHCb \cite{Aaij:2014caa,Aaij:2014b}. Most models in literature overestimate the masses of $\chi_{b1}\left(3P\right)$ and $\chi_{b2}\left(3P\right)$ states. For the  $\chi_{b1}\left(3P\right)$ ref \cite{Wang:2018} employing modified GI model predicts 10527 MeV, the RFT model in ref. \cite{Chen:2019uzm} predicts 10557 MeV, the RQM model in ref. \cite{Godfrey:2015} predicts 10538 MeV. The present work also overestimates the mass by 55 MeV. For the $\chi_{b2}\left(3P\right)$ ref \cite{Wang:2018} employing modified GI model predicts 10539 MeV, the RFT model in ref. \cite{Chen:2019uzm} predicts 10556 MeV, the RQM model in ref. \cite{Godfrey:2015} predicts 10550 MeV.  Our calculated mass is overestimated by about 52 MeV compared to the experimental mass\cite{Zyla:2020zbs}. Also, as the 3P states are above threshold the masses may be affected by coupled channel effect. More theoretical and experimental efforts are desirable for the 3P bottomonium states in future.

\subsubsection{Masses of D states}

For $1^{3}D_{2}$ ($\Upsilon(1D)$) state our calculated mass differs from the experimental mass by 14 MeV. As mentioned earlier in introduction about the newly observed state ${{\mathit \Upsilon}{(10753)}}$ in 2019 at Belle\cite{Abdesselam:2019gth} there is still uncertainty over its structure. Based on the present study we find the mass of $3^{3}D_{1}$ state as 10736 MeV whereas the experimental determined mass for ${{\mathit \Upsilon}{(10753)}}$ is 10753 $\pm$ 6 MeV, the difference being only of 17 MeV. Hence, we assign ${{\mathit \Upsilon}{(10753)}}$ as $3^{3}D_{1}$ state. Same assignment for ${{\mathit \Upsilon}{(10753)}}$ has been made by \cite{Chen:2019uzm}. ${{\mathit \Upsilon}{(10753)}}$  as an admixture of S and D states was ruled out in relativistic flux tube model\cite{Chen:2019uzm} because the di-electron decay width for admixture state can increase by 2 orders. Also, the experimental total decay width calculated is around 36 MeV~\cite{Zyla:2020zbs} which is nearby the total decay width for $3^{3}D_{1}$ state, 54 MeV as calculated by \cite{Wang:2018} which supports our assignment.

\subsection{Decay Constants}
The calculated vector decay constants  $f_{V}$($f_{Vcor}$) and pseudoscalar decay constants $f_{P}$($f_{Pcor}$), without and with QCD corrections, are shown in Table~\ref{tab:decaybb}.The vector decay constants $f_{V}$($f_{Vcor}$) for $1^3S_1$, $2^3S_1$, $3^3S_1$ and $4^3S_1$ states are 65 Mev, 115 MeV, 92 MeV and 17 MeV lower respectively as compared to experimental estimates. Our calculated vector decay constants are in good agreement in comparison to Ref.~\cite{Wang:2006}. Our calculated pseudoscalar decay constants $f_{P}$($f_{Pcor}$) are lower compared to available theoretical estimates.  The $f_{Vcor}$  and $f_{Pcor}$ decay constants are underestimated, where as $f_{V}$ and $f_{P}$  decay constants are overestimated in comparison to the estimates by other available theoretical models.

 \subsection{Radiative Transitions}
The Electrical dipole (E1) transition widths $\varGamma[2S \rightarrow (1P)\gamma]$,  calculated using our predicted bottomonium masses    are shown in Table~\ref{tab:E1BB}. Our results for the E1 transitions width $\varGamma[2^{3}S_1 \rightarrow (1^{3}P_3)\gamma]$, $\varGamma[2^{3}S_1 \rightarrow (1^{3}P_1)\gamma]$ and $\varGamma[2^{3}S_1 \rightarrow (1^{3}P_0)\gamma]$ are 0.51 keV, 0.72 keV and 0.36 keV lower respectively  than the experimental results. Our calculated E1 transition widths $\varGamma[2S \rightarrow (1P)\gamma]$ are comparatively in good agreements with other theoretical estimates.

We have calculated the Electrical dipole (E1) transition widths $\varGamma[1P \rightarrow (1S)\gamma]$, $\varGamma[1D \rightarrow (1P)\gamma]$, $\varGamma[2P \rightarrow (2S)\gamma]$,  $\varGamma[2P \rightarrow (1D)\gamma]$, which are summarized in Table~\ref{tab:E1BB}. Our calculated E1 transitions for  $\varGamma[1^{3}P_2 \rightarrow (1^{3}S_1)\gamma]$ and  $\varGamma[1^{3}P_1 \rightarrow (1^{3}S_1)\gamma]$ are 5.61 keV and 4.26 keV higher respectively as compared to experimental measurements. The E1 transition width $\varGamma[1^{1}P_1 \rightarrow (1^{1}S_0)\gamma]$ is 11.09 keV lower when compared to experimental measurement.

Our calculated E1 transitions for  $\varGamma[2^{3}P_2 \rightarrow (2^{3}S_1)\gamma]$,  $\varGamma[2^{3}P_1 \rightarrow (2^{3}S_1)\gamma]$ and  $\varGamma[2^{1}P_1 \rightarrow (2^{1}S_0)\gamma]$ are 4.62 keV, 5.25 keV and 18.23 keV lower respectively when compared to experimental measurements. Whereas transitions for $\varGamma[2^{3}P_0 \rightarrow (2^{3}S_1)\gamma]$ is much higher but comparable to experimental measurement. Overall our calculated E1 transitions for  $\varGamma[1P \rightarrow (1S)\gamma]$ and $\varGamma[2P \rightarrow (2S)\gamma]$ are comparatively in good agreement with results of other theoretical estimates.  Our calculated E1 transition widths for  $\varGamma[1D \rightarrow (1P)\gamma]$ and $\varGamma[2P \rightarrow (1D)\gamma]$ are in close agreement with the estimates from other theoretical models.

The M1 radiative transition $\varGamma[\Upsilon\left(nS\right)\rightarrow\eta_{b}\left(nS\right)] $ is the unique electromagnetic decay and until now no experimental information is available. Our calculated M1 transition widths using bottomonium masses predicted by our potential model for the low-lying 1S, 2S and 3S states are shown in Table~\ref{tab:M1BB}. Our predicted M1 transition widths are compared with the available predictions from other theoretical models. Our calculated M1 transition width for $\varGamma[1^{3}S_1 \rightarrow (1^{1}S_0)\gamma]$,  $\varGamma[2^{3}S_1 \rightarrow (2^{1}S_0)\gamma]$ and $\varGamma[3^{3}S_1 \rightarrow (3^{1}S_0)\gamma]$ are in close agreement with the estimates by  theoretical models Ref.~\cite{Wang:2018,Segovia2016,Akbar:2015,Godfrey:2015}, while M1 transition width for  $\varGamma[2^{3}S_1 \rightarrow (1^{1}S_0)\gamma]$ and  $\varGamma[2^{1}S_0 \rightarrow (1^{3}S_1)\gamma]$ is higher than the estimates of other theoretical models. But the calculated M1 transition width for $\varGamma[2^{3}S_1 \rightarrow (1^{1}S_0)\gamma]$ is  lower than the experimental M1 transiton width $12.5 \pm  4.9$ keV \cite{Zyla:2020zbs}.

We observe that radiative transitions are model dependent. In practice radiative QCD corrections, play more critical role to the estimation of electromagentic transitions. However estimation of radiative corrections is a separate study in itself and lies beyond the scope of the present work.

 \subsection{Annilation Decays}

We calculate the partial decay widths $\varGamma$ and $\varGamma^{cf}$ (with QCD correction factor) of annihilation processes for $e^{+}e^{-}$, di-photon, tri-photon, di-gluon, tri-gluon, $\gamma gg$ and $q\bar{q}+g$. The results are  compared with experimental data from the PDG~\cite{Zyla:2020zbs} as well as  with other theoretical estimates.

For the bottomonium states $\Upsilon(nS)$, we calculate leptonic decay widths $\varGamma_{l^{+}l^{-}}$ and $\varGamma_{l^{+}l^{-}}^{cf}$ in keV, tabulated in Table~\ref{annihi2e}, tri-photon decay widths  $\varGamma_{3\gamma}$ and $\varGamma_{3\gamma}^{cf}$ in eV, tabulated in Table~\ref{annihi3p}, tri-gluon decay widths $\varGamma_{3g}$ and $\varGamma_{3g}^{cf}$ in keV, tabulated in Table~\ref{annihi3g} as well as $n^{3}S_1\rightarrow \gamma gg$ decay widths in keV, tabulated in Tablet~\ref{annihip2g}.
Our estimated leptonic widths without and with QCD correction are lower compared to experimentally observed widths as well as other theoretically estimates from Ref.~\cite{Godfrey:2015,Wang:2018,Li:2009nr,Bhaghyesh:2011}, whereas calculated leptonic widths are in good agreement with estimates from Ref.~\cite{Segovia2016,Chaturvedi:2020,Gonzalez:2004}.
Our calculated tri-photon and tri-gluon decay widths are in fine tune with available theoretical estimates. Our calculated $n^3S_1\rightarrow\gamma{gg}$ decay widths are higher when compared to experimental data but are comparable to the widths calculated by other theoretical models.

 For the  $\eta_{b}(nS)$  states, we calculate di-photon decay width $\varGamma_{\gamma\gamma}$ and $\varGamma_{\gamma\gamma}^{cf}$ in keV, as well as di-gluon decay widths $\varGamma_{gg}$ and $\varGamma_{gg}^{cf}$ in MeV and the results are summarized in Table~\ref{annihi2p} and \ref{annihi2g} respectively. The calculated di-photon decay widths for 1S and 2S states agree well with the experimentally determined decay widths when calculated excluding the first order QCD radiative correction, while the decay width drops when the first order QCD radiative correction is added. When comparing the calculated decay width with the results from other theoretical approaches our results are in accordance with the decay widths of Ref. \cite{Munz:1996,Ebert2003a,Gupta:1996,Li:2009nr}. The decay widths of Ref. \cite{Wang:2018,Segovia2016,Godfrey:2015,Anisovich:2005} are very high and not even closer to the experimental determined results, thus this study favors our calculated decay widths. The decay widths of all the P wave states are very low and no experimental result is available for a bold statement.\\

We have calculated leptonic decay widths $n^{3}D_1\rightarrow\varGamma_{l^{+}l^{-}}$, $\varGamma_{l^{+}l^{-}}^{cf}$ in eV and results are shown in Table~\ref{annihi2e}. Our calculated leptonic decay widths without QCD corrections for $n^{3}D_1$ are higher compared to leptonic decay width estimated by other theoretical models. After applying QCD correction, leptonic decay widths for $n^{3}D_1$ are comparatively in good agreement with estimated by available theoretical models.

Our calculated di-photon decay widths for $n^{3}P_0\rightarrow\varGamma_{\gamma\gamma}$, $\varGamma_{\gamma\gamma}^{cf}$ and $n^{3}P_{2}\rightarrow\varGamma_{\gamma\gamma}$, $\varGamma_{\gamma\gamma}^{cf}$ in keV are shown Table~\ref{annihi2p}. We have compared our di-photon decay widths with the widths from available theoretical models. we observe that our calculated di-photon decay width results for $n^{3}P_0$ states are lower when compared to the decay widths predicted by other theoretical models, whereas decay widths for $n^{3}P_{2}$ states are lower in comparison to decay widths predicted by Ref.~\cite{Godfrey:2015,Wang:2018,Chaturvedi:2020,Anisovich:2005,Segovia2016,Li:2009nr,Laverty:2009,Munz:1996}.

Our calculated di-gluon decay widths $n^{3}P_{0}\rightarrow\varGamma_{gg}$, $\varGamma_{gg}^{cf}$ in MeV,  $n^{3}P_{2}\rightarrow\varGamma_{gg}$, $\varGamma_{gg}^{cf}$ in MeV and $n^{1}D_{2}\rightarrow\varGamma_{gg}$, $\varGamma_{gg}^{cf}$ in keV are shown Table~\ref{annihi2g}.
For $n^{3}P_0$ states our calculated di-gluon decay widths (with and without QCD correction) are lower compared to decay widths predicted by Ref.~\cite{Wang:2018,Segovia2016,Godfrey:2015} while it is comparatively in good agreements with decay widths predicted by Ref.~\cite{Pandya:2021,Laverty:2009,Ebert2003}. Our calculated di-gluon decay widths for the $n^{3}P_2$ and  $n^{1}D_2$ states, are in good agreements with the other theoretical estimates.

We also calculate the tri-gluon decay widths for $n^{1}P_1$, $n^{3}D_1$, $n^{3}D_2$ and $n^{3}D_3$ states and results are tabulated in Table~\ref{annihi3g}. Tri-gluon decay widths for $n^{1}P_1$ state are a little bit on the lower side. Tri-gluon decay widths for $n^{3}D_1$, $n^{3}D_2$ and $n^{3}D_3$ states when compared with other available theoretical estimates seems to be in good agreement. We also calculate $n^3P_1\rightarrow q\bar{q}+g$ decay width and results are tabulated in Table~\ref{annihiqqg}. Our calculated decay widths are comparatively low with respect to Ref.~\cite{Wang:2018,Segovia2016,Godfrey:2015} and in good agreements with Ref.~\cite{Ebert2003,Pandya:2021}.

 We also observe the wide variation in estimated values of annihilation decay width, may be due to  different treatments in the relativistic corrections and the considerable uncertainties which arise from the wave function dependence of the model.

 \subsection{Regge trajectories \label{sec:reg}}

We construct the Regge trajectories for the $(n,M^{2})$ and  $(J,M^{2})$ planes using bottomonium masses estimated by our potential model. The trajectories with the same value of $J$ and differ by a quantum number corresponding to the radial quantum number are usually called "daughter" trajectories and their masses are higher than those of the leading trajectory with given quantum numbers. The linearity of Regge trajectories is interpreted as a expression of strong forces between quarks at large distances(color confinement). The Regge trajectories are basically nonlinear for mesons with massive quarks \cite{Kruczenski:2004}.

\begin{figure}
        \centering
        \includegraphics[bb=30bp 60bp 750bp 550bp,clip,width=0.80\textwidth]{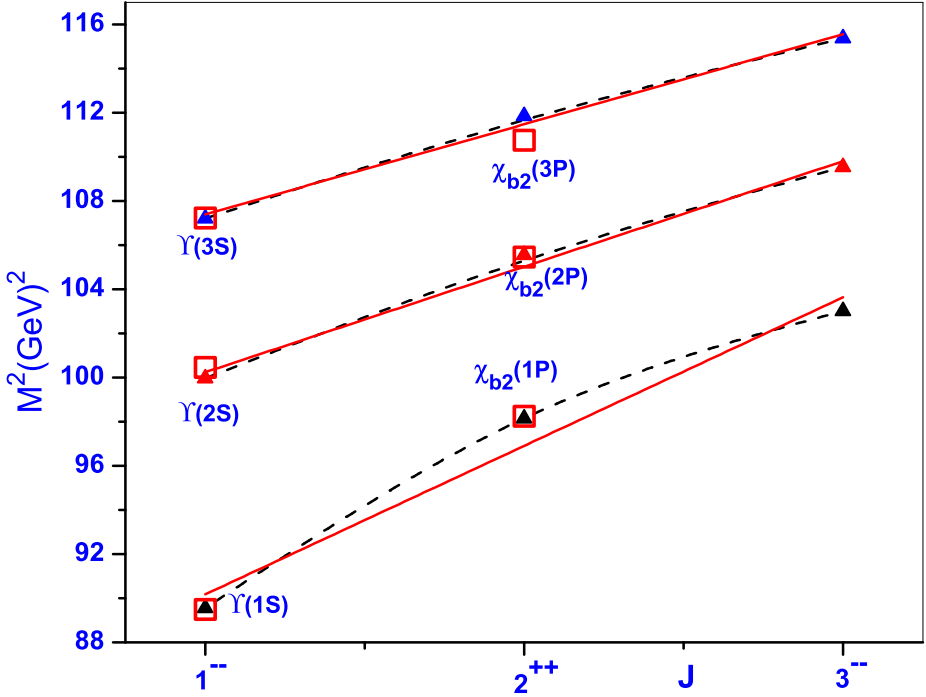}
        \caption{Regge trajectory ($ J \rightarrow M^{2}$) with natural parity. \label{fig:NPmesonBB}}
   \end{figure}

\begin{figure}
      \centering
      \includegraphics[bb=30bp 60bp 750bp 550bp,clip,width=0.80\textwidth]{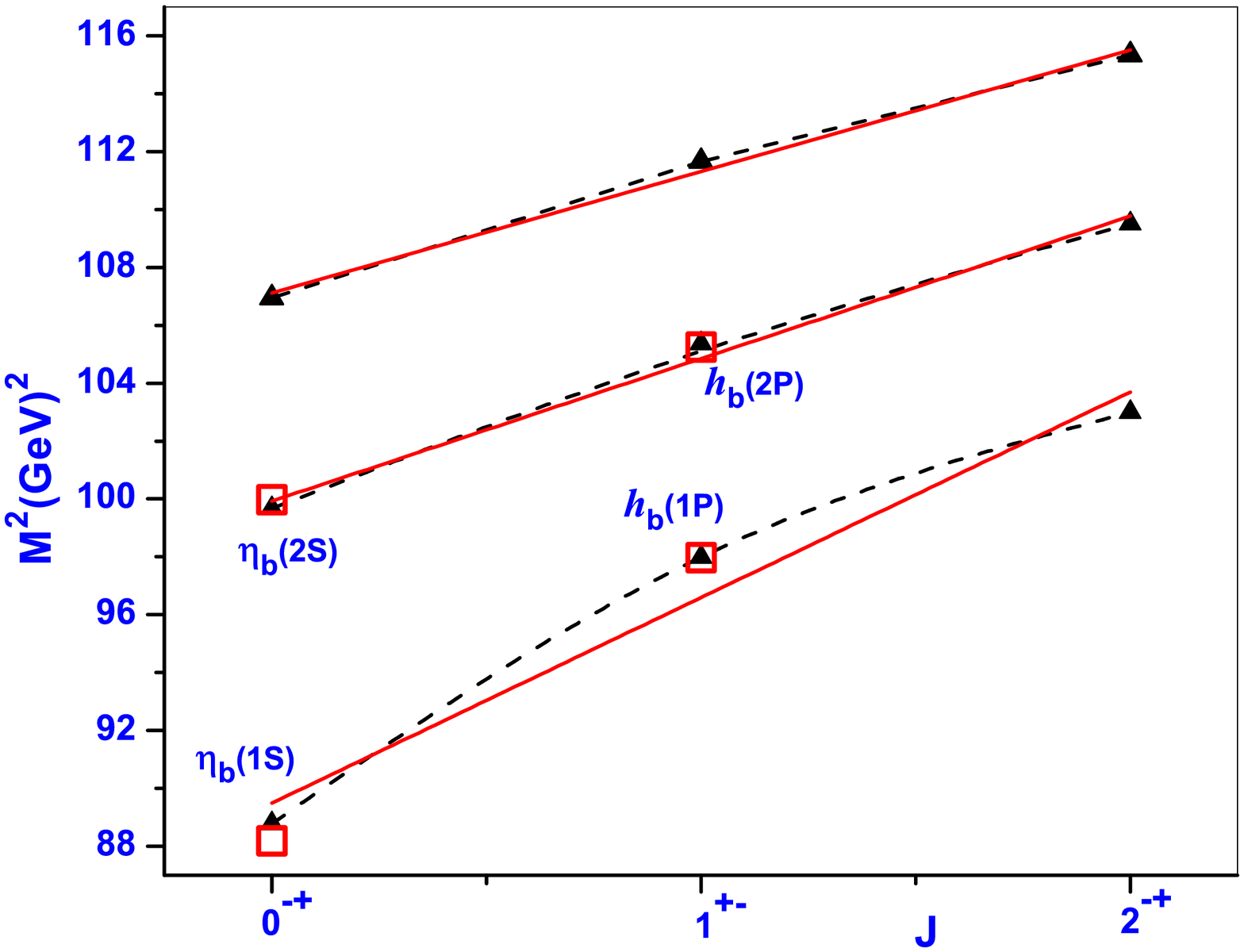}
    \caption{Regge trajectory ($J \rightarrow M^{2}$) with unnatural parity. \label{fig:UNPmesonBB}}
     \end{figure}

\begin{figure}
     \centering
      \includegraphics[bb=30bp 60bp 750bp 550bp,clip,width=0.80\textwidth]{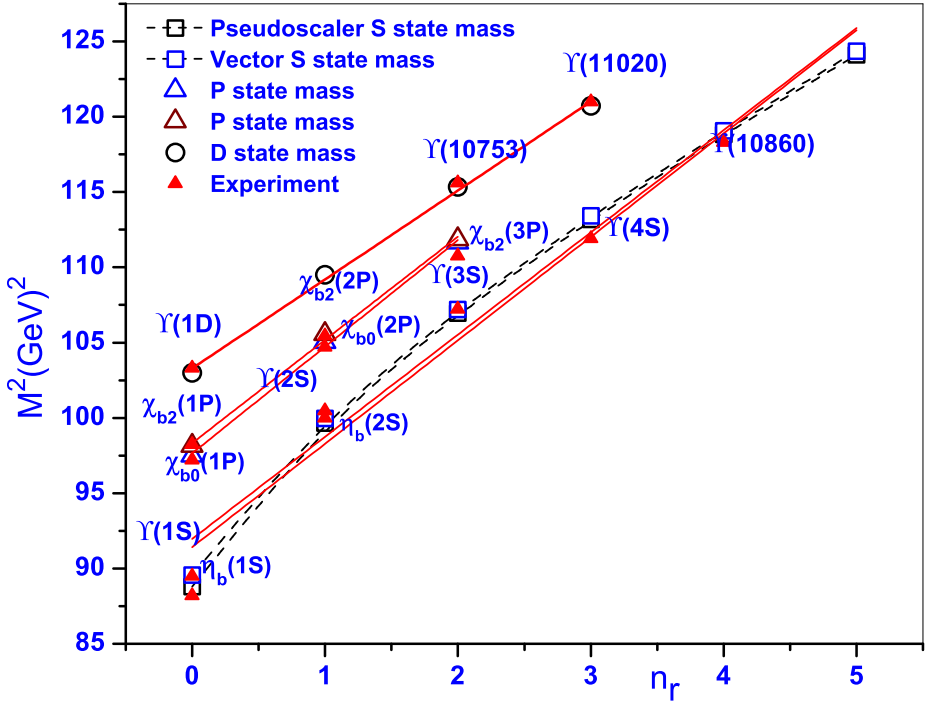}
      \caption{Regge trajectory ($n_{r}\rightarrow M^{2}$) for the pseudoscalar and vector $S$ state and excited $P$ and $D$ state masses.\label{fig:PsVmesonBB}}
     \end{figure}

\begin{figure}
      \centering
      \includegraphics[bb=30bp 60bp 750bp 550bp,clip,width=0.80\textwidth]{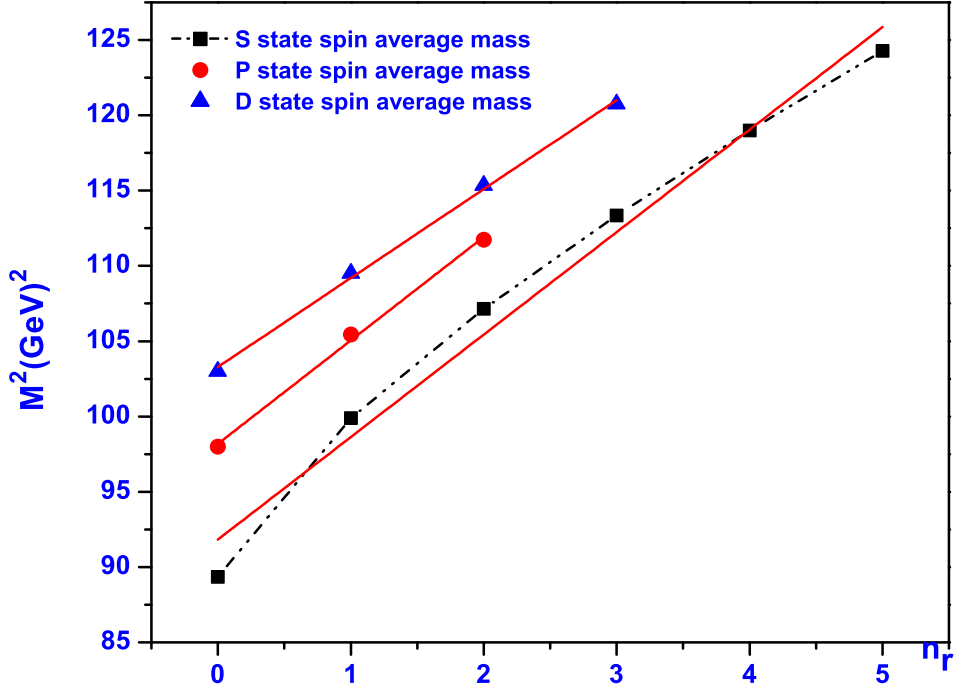}
      \caption{Regge trajectory ($ n_{r} \rightarrow M^{2}$) for the S-P-D states center of weight mass.\label{fig:SavmesonBB}}
      \end{figure}

The Regge trajectories in the $(J,M^{2})$ plane with $(P=(-1)^{J})$ ($J^{P}=1^{-},2^{+},3^{-}$) natural and $(P=(-1)^{J-1})$ ($J^{P}=0^{-},1^{+},2^{-}$) unnatural parity are plotted in Figs.~\ref{fig:NPmesonBB}-\ref{fig:UNPmesonBB}. In the diagrams, the solid triangles represent the bottomonium masses estimated by our potential model  and the hollow squares represent experimentally available masses with the corresponding bottomonium name.  The Regge trajectories for principal quantum number $n_{r}= n-1$ in the $(n_{r},M^{2})$ plane are shown in Figs.~\ref{fig:PsVmesonBB} and \ref{fig:SavmesonBB}.

 To calculate the $\chi^{2}$  fitted slopes ($\alpha$, $\beta$) and the intercepts ($\alpha_{0}$, $\beta_{0}$), definitions used are \cite{Kher2017b,Kher2017c,kher2018}:
  \begin{equation}
        J=\alpha M^{2}+\alpha_{0}.\label{eq:J regge}
  \end{equation}
 \begin{equation}
        n_{r}=\beta M^{2}+\beta_{0}\label{eq:nr regge}
  \end{equation}

The calculated slopes and intercepts are tabulated in Tables~\ref{tab:alfaBB},~\ref{tab:bitaBB}, and~\ref{tab:SpinaveBB}). The calculated masses of  the bottomonium fit well into the $(n,M^{2})$ and  $(J,M^{2})$ planes. The parent Regge trajectories in both ($J, M^2$)  and ($n_r, M^2$) planes, which start from ground  states, are exhibiting a nonlinear behavior in the lower mass region, whereas the daughter trajectories, which involve both  radially and orbitally excited states, turn out to be almost linear, equidistant and parallel. The slopes of Regge trajectories in $(J, M^{2})$ and ($n_r, M^2$) planes, increases with increasing quark masses.
The  orbital momentum $\ell$ of the state is proportional to its mass: $\ell=\alpha M^{2}(\ell)+\alpha(0)$, where the slope $\alpha$ depends on the flavor content of the states, hence quark masses affect the linearity of the Regge trajectories. In the framework of the hadron string model,the radial spectrum of heavy quarkonia typically leads to strong nonlinearities \cite{Afonin2016}.

  \begin{table*}
    \caption{Slopes and intercepts of the $(J,\: M^{2})$ Regge trajectories with unnatural and natural parity.\label{tab:alfaBB} }

    \noindent \centering{}%
           \begin{tabular}{cccc}
   \hline
  \addlinespace[3pt]
   {Parity}  & {Trajectory} & {$\alpha(GeV^{-2})$} & {$\alpha_{0}$}\tabularnewline
    \addlinespace[2pt]
  \hline
   \addlinespace[3pt]
         \multirow{3}{*}{Unnatural } & Parent & $0.234\pm0.023$ & $-6.587\pm0.741$\tabularnewline
           \smallskip
          & First daughter & $0.244\pm0.016$ & $-8.616\pm0.634$\tabularnewline
            \smallskip
          & Second daughter & $0.235\pm0.006$ & $-9.913\pm0.261$\tabularnewline
   \addlinespace[2pt]
  \hline
   \addlinespace[3pt]
    \multirow{3}{*}{Natural}  & Parent &  $0.145\pm0.023$ & $-12.036\pm2.241$\tabularnewline
             \smallskip
         & First daughter &  $0.207\pm0.020$ & $-19.754\pm2.105$\tabularnewline
             \smallskip
         & Second daughter & $0.243\pm0.019$ & $-25.139\pm2.152$\tabularnewline
   \addlinespace[2pt]
  \hline
  \end{tabular}
  \end{table*}
  \begin{table}
  \caption{Slopes and intercepts of the $(n_{r}, M^{2})$ Regge trajectories. \label{tab:bitaBB}}
   \noindent \begin{centering}
        \begin{tabular}{cccc}
  \hline
   \addlinespace[3pt]
            Meson & $J^P$ & $\beta(GeV^{-2})$ & $\beta_{0}$\tabularnewline
   \addlinespace[2pt]
  \hline
   \addlinespace[3pt]
  $\eta_{b}$ & $0^{-+}$ & $0.133\pm0.011$ & $-12.061\pm1.133$\tabularnewline
  $\Upsilon$ & $1^{--}$ & $0.136\pm0.010$ & $-12.367\pm1.073$\tabularnewline
  $\chi_{b0}$ & $0^{++}$ & $0.141\pm0.005$ & $-13.780\pm0.543$\tabularnewline
  $\chi_{b1}$ & $1^{++}$ & $0.146\pm0.007$ & $-14.342\pm0.697$\tabularnewline
  $h_{b}$ & $1^{+-}$ & $0.144\pm0.007$ & $-14.149\pm0.784$\tabularnewline
  $\chi_{b2}$ & $2^{++}$ & $0.146\pm0.007$ & $-14.314\pm0.724$\tabularnewline
  $\Upsilon(^{3}D_{1})$ & $1^{--}$ & $0.162\pm0.005$ & $-16.674\pm0.578$\tabularnewline
  $\Upsilon(^{3}D_{2})$ & $2^{--}$ & $0.163\pm0.006$ & $-16.780\pm0.641$\tabularnewline
  $\Upsilon(^{1}D_{2})$ & $2^{-+}$ & $0.162\pm0.005$ & $-16.683\pm0.574$\tabularnewline
  $\Upsilon(^{3}D_{3})$ & $3^{--}$ & $0.162\pm0.005$ & $-16.679\pm0.570$\tabularnewline
   \addlinespace[2pt]
  \hline
  \end{tabular}
    \par\end{centering}
    \end{table}

   \begin{table}
   \caption{Slopes and intercepts of  $(n_{r}, M^{2})$ Regge trajectory for center of weight mass.\label{tab:SpinaveBB} }
    \noindent \centering{}%
          \begin{tabular}{ccc}
    \hline
     \addlinespace[3pt]
           Trajectory & $\beta(GeV^{-2})$ & $\beta_{0}$\tabularnewline
        \addlinespace[2pt]
       \hline
        \addlinespace[3pt]
   S State & $0.144\pm0.010$ & $-13.216\pm1.049$ \tabularnewline
   P State & $0.145\pm0.007$ & $-14.25\pm0.737$ \tabularnewline
   D State & $0.162\pm0.005$ & $-16.686\pm0.570$ \tabularnewline
   \addlinespace[2pt]
  \hline
 \end{tabular}
   \end{table}


\section{Conclusion\label{sec:conclusion}}

Our predicted masses for the $S-P-D$ states from the present study are very close to the available experimental as well as the other theoretical estimates. The Regge trajectories in both ($J, M^2$) and ($n_r, M^2$) planes are almost linear, equidistant and parallel. From Figs. \ref{fig:NPmesonBB}-\ref{fig:UNPmesonBB}, we find that the experimental states are  sitting nicely on straight lines without deviation.  The pseudoscalar $(f_{Pcor})$ and the vector $(f_{Vcor})$ decay constants with QCD correction are close to available theoretical predictions. The predicted radiative (E1 and M1 dipole) transition widths are in satisfactory agreement with theoretical and experimental predictions. From comparison of our estimated radiative transitions widths with other theoretical estimations, we observed that due to different treatments or parameters in the relativistic corrections in the model, the various models have very different predictions. Although more precise experimental measurements are required in most cases.
Using the Van Royen-Weisskopf relation, we predict annihilation Decay widths and we conclude that the inclusion of QCD correction factors are of importance for obtaining accurate results. The results of annihilation Decay widths in various models again show a wide range of variations. \\

{\bf Acknowledgements} A. K. Rai acknowledges the financial support extended by the Department of Science of Technology, India  under the SERB fast track scheme SR/FTP /PS-152/2012.\\

\bibliographystyle{spphys}
\bibliography{BB_meson}

\end{document}